\documentclass[journal]{IEEEtran}

\usepackage[T1]{fontenc}
\usepackage[utf8]{inputenc}
\usepackage[english]{babel}

\usepackage{amsmath, amssymb, amsfonts, bm}

\usepackage{graphicx}
\usepackage{subcaption} % For subfigures (modern and recommended)
\usepackage{tikz}
\usetikzlibrary{matrix, calc, arrows.meta, positioning, angles, quotes, decorations.pathmorphing, shapes.geometric}

\usepackage{pgfplots}
\usepackage{pgfplotstable}
\usepgfplotslibrary{fillbetween} % Útil se precisar preencher áreas no futuro
\pgfplotsset{compat=1.18} % Garante que a formatação e âncoras usem regras modernas

\usepackage{cite}

\usepackage{url}
\usepackage{adjustbox}
\usepackage{booktabs}       % For high-quality tables
\usepackage{algorithm}      % For algorithm environment
\usepackage{algpseudocode}  % For Cormen-style pseudocode
\usepackage{siunitx}        % For SI units (e.g., \SI{10}{\hertz})
\usepackage{empheq}         % For highlighting equations (e.g., with boxes)
\usepackage{footnote}       % Better control of footnotes
\usepackage{microtype}      % Improves text spacing and justification
\usepackage{setspace}       % For line spacing control
\usepackage{enumitem}       % For customizing lists
\usepackage{soul}           % For underlining, highlighting, etc. (use sparingly)
\usepackage{comment}        % For comment blocks
\usepackage{tabularx}
\usepackage{makecell}
\usepackage{fontawesome5}
\usepackage[acronym,nonumberlist]{glossaries}
\usepackage{stfloats}
\usepackage[colorlinks=true, linkcolor=blue, citecolor=red, urlcolor=cyan, bookmarks=false]{hyperref}

\newenvironment{anup}
  {\begingroup\color{black}}
  {\endgroup}
\newacronym{3gpp}{3GPP}{3rd generation partnership project}
\newacronym{5g}{5G}{5th generation}
\newacronym{6g}{6G}{sixth-generation}
\newacronym{ack}{ACK}{acknowledgement}
\newacronym{afdm}{AFDM}{affine frequency division multiplexing}
\newacronym{aoi}{AoI}{age-of-information}
\newacronym{ap}{AP}{access point}
\newacronym{awgn}{AWGN}{additive white Gaussian noise}
\newacronym{b5g}{B5G}{beyond 5G}
\newacronym{ber}{BER}{bit-error rate}
\newacronym{bp}{BP}{belief propagation}
\newacronym{bs}{BS}{base station}
\newacronym{cscg}{CSCG}{circularly symmetric complex Gaussian}
\newacronym{csi}{CSI}{channel state information}
\newacronym{dd}{DD}{delay--doppler}
\newacronym{dir}{DIR}{data information rate}
\newacronym{dof}{DoF}{degrees-of-freedom}
\newacronym{drl}{DRL}{deep rl}
\newacronym{dtmc}{DTMC}{discrete-time markov chain}
\newacronym{embb}{eMBB}{enhanced mobile broadband}
\newacronym{fdma}{FDMA}{frequency division multiple access}
\newacronym{fec}{FEC}{forward error correction}
\newacronym{fifo}{FIFO}{first-in first out}
\newacronym{fim}{FIM}{Fisher information matrix}
\newacronym{ga}{GA}{genetic algorithm}
\newacronym{gap}{GAP}{generalized assignment problem}
\newacronym{geo}{GEO}{geosynchronous earth orbit}
\newacronym{gsl}{GSL}{ground-to-satellite link}
\newacronym{icsi}{ICSI}{imperfect channel state information}
\newacronym{irsa}{IRSA}{irregular repetition slotted aloha}
\newacronym{isa}{ISA}{isolated spectral allocation}
\newacronym{isac}{ISAC}{integrated sensing and communication}
\newacronym{isic}{ISIC}{imperfect successive interference cancellation}
\newacronym{isl}{ISL}{inter-satellite link}
\newacronym{iot}{IoT}{internet of things}
\newacronym{leo}{LEO}{low earth orbit}
\newacronym{lmmse}{LMMSE}{linear minimum mean square error}
\newacronym{los}{LoS}{line-of-sight}
\newacronym{mab}{MAB}{multi-armed bandit}
\newacronym{mdp}{MDP}{markov decision process}
\newacronym{mgap}{MGAP}{multi-level generalized assignment problem}
\newacronym{mimo}{MIMO}{multiple-input multiple-output}
\newacronym{miso}{MISO}{multiple-input single-output}
\newacronym{mmimo}{mMIMO}{massive mimo}
\newacronym{mmse}{MMSE}{minimum mean square error}
\newacronym{mmtc}{mMTC}{massive machine-type communications}
\newacronym{mp}{MP}{message passing}
\newacronym{mu}{MU}{multi-user}
\newacronym{nack}{NACK}{negative ack}
\newacronym{noma}{NOMA}{non-orthogonal multiple access}
\newacronym{nr}{NR}{new radio}
\newacronym{ntn}{NTN}{non-terrestrial network}
\newacronym{ofdm}{OFDM}{orthogonal frequency division multiplexing}
\newacronym{ofdma}{OFDMA}{orthogonal frequency-division multiple access}
\newacronym{oma}{OMA}{orthogonal multiple access}
\newacronym{otfs}{OTFS}{orthogonal time frequency space}
\newacronym{pdf}{PDF}{probability density function}
\newacronym{pwf}{PWF}{phase warping function}
\newacronym{qos}{QoS}{quality of service}
\newacronym{ran}{RAN}{radio access network}
\newacronym{reir}{REIR}{radar estimation information rate}
\newacronym{rl}{RL}{reinforcement learning}
\newacronym{rms}{RMS}{root mean square}
\newacronym{rs}{RS}{rate-splitting}
\newacronym{rsma}{RSMA}{rate-splitting multiple access}
\newacronym{sagin}{SAGIN}{space--air--ground integrated network}
\newacronym{sc}{SC}{superposition coding}
\newacronym{sdma}{SDMA}{space-division multiple access}
\newacronym{sic}{SIC}{successive interference cancellation}
\newacronym{sinr}{SINR}{signal-to-interference-plus-noise ratio}
\newacronym{siso}{SISO}{single-input single-output}
\newacronym{snr}{SNR}{signal-to-noise ratio}
\newacronym{sp}{SP}{signal processing}
\newacronym{tacae}{TCAE}{time-averaged cost of actuation error}
\newacronym{tare}{TRE}{time-averaged reconstruction error}
\newacronym{tgp}{TGP}{tunable Gaussian pulse}
\newacronym{tdma}{TDMA}{time division multiple access}
\newacronym{tf}{TF}{time--frequency}
\newacronym{uav}{UAV}{unmanned aerial vehicle}
\newacronym{uc}{UC}{update-delivery cost}
\newacronym{ue}{UE}{user equipment}
\newacronym{urllc}{URLLC}{ultra-reliable and low-latency communications}
\newacronym{v2x}{V$2$X}{vehicle-to-everything}

\newacronym{af}{AF}{ambiguity function}
\newacronym{ftn}{FTN}{faster-than-Nyquist}
\newacronym{csit}{CSIT}{channel state information at the transmitter}
\newacronym{csir}{CSIR}{channel state information at the receiver}
\newacronym{acf}{ACF}{auto-correlation function}
\newacronym{crlb}{CRLB}{Cram\'er-Rao lower bound}
\newacronym{rrc}{RRC}{root raised cosine}
\newacronym{ul}{UL}{uplink}
\newacronym{semi-isac}{Semi-ISaC}{semi-integrated sensing and communication}
\newacronym{cu}{CU}{communication user}
\newacronym{rt}{RT}{radar target}
\newacronym{rcs}{RCS}{radar cross-section}
\newacronym{dl}{DL}{downlink}
\newacronym{op}{OP}{outage probability}
\newacronym{wssus}{WSSUS}{wide-sense stationary uncorrelated scattering}
\newacronym{sgp}{SGP}{standard Gaussian pulse}
\newacronym{gbps}{Gbps}{gigabits per second}
\newacronym{oobe}{OOBE}{out-of-band emission}
\newacronym{papr}{PAPR}{peak-to-average power ratio}
\newacronym{ipr}{IPR}{interference power ratio}
\newacronym{ici}{ICI}{inter-carrier interference}
\newacronym{isi}{ISI}{inter-symbol interference}
\newacronym{rr}{RR}{radar receiver}
\newacronym{dmrs}{DMRS}{demodulation reference signals}
\newacronym{pusch}{PUSCH}{physical uplink shared channel}
\newacronym{cp}{CP}{cyclic prefix}
\newacronym{ifft}{IFFT}{inverse fast Fourier transform}
\newacronym{fft}{FFT}{fast Fourier transform}
\newacronym{cfr}{CFR}{channel frequency response}
\newacronym{dft}{DFT}{discrete Fourier transform}
\newacronym{dpi}{DPI}{direct-path interference}
\newacronym{tx}{TX}{transmitter}
\newacronym{1tfde}{1T-FDE}{one-tap frequency-domain equalization}
\newacronym{ta}{TA}{timing advance}
\newacronym{wmmse}{WMMSE}{weighted minimum mean square error}
\newacronym{se}{SE}{spectral efficiency}
\newacronym{dc}{DC}{difference-of-convex}
\newacronym{sca}{SCA}{successive convex approximation}
\newacronym{nphard}{NP-hard}{non-deterministic polynomial-time hard}
\newacronym{lmi}{LMI}{linear matrix inequality}
\newacronym{soc}{SOC}{second-order cone}
\newacronym{sqp}{SQP}{sequential quadratic programming}
\newacronym{kpi}{KPI}{key performance indicator}
\newacronym{fr2}{FR2}{frequency range 2}
\newacronym{fdm}{FDM}{frequency division multiplexing}
\newacronym{rmse}{RMSE}{root mean square error}
\newacronym{kkt}{KKT}{Karush-Kuhn-Tucker}
\newacronym{tdl}{TDL}{tapped delay line}
\newacronym{adc}{ADC}{analog-to-digital converter}
\newacronym{isd}{ISD}{inter-site distance}
\newacronym{v2n}{V2N}{vehicle-to-network}
\newacronym{upa}{UPA}{uniform planar array}
\newacronym{uma}{UMa}{urban macro}
\newacronym{nlos}{NLOS}{non-line-of-sight}
\newacronym{cpi}{CPI}{coherent processing interval}
\newacronym{mmwave}{mmWave}{millimeter wave}
\newacronym{idft}{IDFT}{inverse discrete Fourier transform}
\newacronym{td}{TD}{time-domain}
\newacronym{fd}{FD}{frequency-domain}
\newacronym{dp}{DP}{direct-path}
\newacronym{ep}{EP}{echo-path}
\newacronym{mrc}{MRC}{maximum-ratio combining}
\newacronym{mse}{MSE}{mean square error}
\newacronym{bcd}{BCD}{block coordinate descent}
\newacronym{fpp}{FPP}{Feasible Point Pursuit}
\newacronym{qcqp}{QCQPs}{Quadratically constrained quadratic programs}
\newacronym{fp}{FP}{fractional programming}
\newacronym{sdp}{SDP}{semidefinite program}
\newacronym{ao}{AO}{alternating optimization}
\newacronym{psd}{PSD}{positive semi-definite}
\newacronym{ifi}{IFI}{inter-functionality interference}
\newacronym{pslr}{PSLR}{peak sidelobe ratio}
\newacronym{rf}{RF}{radio frequency}
\newacronym{bler}{BLER}{block error rate}

\definecolor{cadmiumgreen}{rgb}{0.0, 0.42, 0.24}
\definecolor{ao(english)}{rgb}{0.0, 0.5, 0.0}
\newcommand{\bc}{\textcolor{black}}         % Bruno's comments
\newcommand{\am}{\textcolor{black}}    % Anup's comments
\newcommand{\ilm}{\textcolor{black}}          % Israel's comments
\newcommand{\ta}{\textcolor{black}}           % Taufik's comments
\newcommand{\pp}{\textcolor{black}}         % Petar's comments
\definecolor{twi_optimal}{HTML}{d7191c} % Vermelho do NOMA-CF
\definecolor{path2}{HTML}{bbbbbb}       % Cinza do NOMA-SF
\definecolor{matlab_blue}{rgb}{0.00, 0.45, 0.74}
\definecolor{radar_yellow}{RGB}{237,177,32} % Amarelo (Radar)

\begin{document}

% ===================================================================
% === TITLE, AUTHORS AND AFFILIATIONS ===============================
% ===================================================================
\title{Rate-Splitting--Inspired Bistatic OFDM-ISAC}

\author{Bruno~F.~Costa,~\IEEEmembership{Student Member,~IEEE,}
        Anup~Mishra,~\IEEEmembership{Member,~IEEE,}
        \ilm{Israel~Leyva-Mayorga},~\IEEEmembership{Member,~IEEE,}
        \ta{Taufik~Abrão},~\IEEEmembership{Senior Member,~IEEE,}
        and~\pp{Petar~Popovski},~\IEEEmembership{Fellow,~IEEE} \vspace{-0.7cm}%
\thanks{B. F. Costa and T. Abrão are with the Department of Electrical Engineering, State University of Londrina (UEL), Londrina, PR, 86057-970, Brazil (e-mail: bruno.felipe.costa@uel.br; taufik@uel.br)}%
\thanks{A. Mishra, I. Leyva-Mayorga, and P. Popovski are with the Department of Electronic Systems, Aalborg University,  Denmark (e-mail: anmi@es.aau.dk; ilm@es.aau.dk; petarp@es.aau.dk).}
}

% ===================================================================
% === ARTICLE HEADERS ===============================================
% ===================================================================
%\markboth{IEEE Transactions on Wireless Communications,~Vol.~XX,~No.~X,~Month~2026}%
%{Costa \MakeLowercase{\textit{et al.}}: Rate-Splitting for OFDM-based ISAC in Doubly-Selective Channels}

\markboth{}%
{Costa \MakeLowercase{\textit{et al.}}: Rate-Splitting--Inspired Bistatic OFDM-ISAC}

% ===================================================================
% === DOCUMENT CONTENT BEGINS =======================================
% ===================================================================
\maketitle
\glsresetall
\begin{abstract}
Achieving effective uplink bistatic \gls{isac} over an \gls{ofdm}
waveform gives rise to challenging interference structures. These are
mostly due to unequal direct- and echo-path contributions and
Doppler-induced \gls{ici}, \bc{which challenge both orthogonal resource
separation and fixed-order \gls{sic} strategies for interference management.}
To address this problem, \bc{we consider a \gls{rs}-inspired approach for
uplink bistatic \gls{ofdm}-\gls{isac}, in which the communication message
is split into robust and supplementary streams that are superposed with a
sensing signal. We develop a staged sensing--communication receiver that
exploits this flexibility to manage interference across the communication
and sensing processing stages.} \bc{The resulting formulation captures the
coupling between sensing uncertainty, echo reconstruction, and post-sensing
communication performance,} and leads to a sensing-constrained
power-allocation problem for communication \gls{se} maximization.
\bc{We develop a tractable iterative solution based on convex approximation
and fractional programming.} \bc{Numerical results show that the \gls{rs}-inspired design provides
effective-\gls{se} gains over fixed-order communication-first and
sensing-first \gls{noma}-inspired baselines, with the 
communication split adapting to the echo-path strength, Doppler-induced
\gls{ici}, and sensing requirement.}
\begin{comment}
Achieving effective uplink bistatic \gls{isac} over an \gls{ofdm}
waveform gives rise to challenging interference structures. These arise
primarily from unequal direct- and echo-path contributions and
Doppler-induced \gls{ici}, which challenge both orthogonal resource
separation and fixed-order \gls{sic} strategies. To address this problem, we propose a \gls{rs}-inspired framework \bc{for a single communication \gls{ue}, where the transmitter splits one communication message into a robust and a supplementary stream, which are jointly superposed over a sensing signal}. \bc{Furthermore, we present the design of a staged sensing and communication receiver}. Based on this framework, \bc{we derive tractable per-subcarrier \gls{sinr} expressions and establish the sensing and communication link through an \gls{sinr}-weighted sensing-information surrogate}. Building on these, we formulate a joint power-allocation problem for \gls{se} maximization under sensing-performance and power constraints. The resulting non-convex formulation is solved using convex surrogates and fractional programming. Numerical results \bc{quantify the spectral-efficiency gain achieved relative to the independently optimized communication-first (\gls{noma}-CF) and sensing-first (\gls{noma}-SF) endpoint architectures and their realization-wise envelope, at zero and nonzero mean target Doppler under a common relative sensing calibration}.
\end{comment}
\end{abstract}


\glsresetall

% ============================= I. INTRODUCTION =============================
\vspace{-0.3cm}
\section{Introduction}
\Gls{isac} has been touted as a key enabling technology for \gls{6g} wireless networks, driven by the growing convergence of spectrum usage, hardware platforms, and signal processing pipelines for communication and sensing \cite{FanLiu2022a,Yuanwei_NOMA_ISaC,Mishra2025,FanLiu2020}. By allowing the same wireless infrastructure to provide both connectivity and environmental awareness, \gls{isac} promises gains in spectral and energy efficiency while enabling applications such as intelligent transportation, remote monitoring, and autonomous systems \cite{FanLiu2022a,mishra2025temporal}. These gains, however, come at the cost of more intricate interaction between communication and sensing, since sharing spectral resources, waveform dimensions, and receiver processing makes mutual interference unavoidable \cite{chen2025interference}. Consequently, interference management emerges as a defining design challenge in \gls{isac}, rather than a secondary implementation issue \cite{chen2025interference,FanLiu2020,Yuanwei_NOMA_ISaC}.
\begin{figure}[!htbp]
    \centering
    \includegraphics[width=\columnwidth]{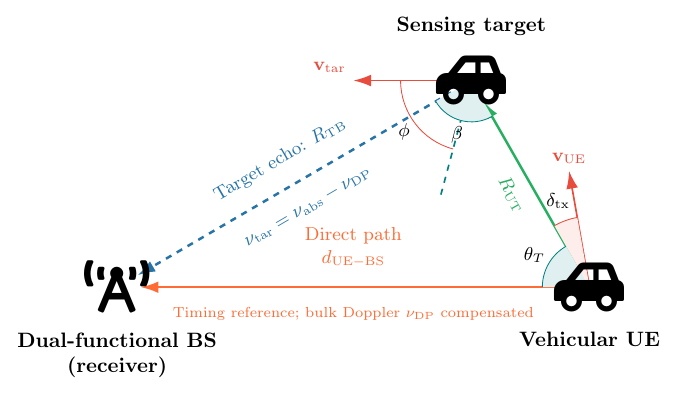}
    \caption{\bc{Uplink bistatic \gls{isac} geometry, shown schematically and not to scale. Velocity vectors and angles $\delta_{\mathrm{tx}}$, $\theta_T$, $\beta$, and $\phi$ specify the Doppler projections; the echo Doppler is referenced to the compensated direct path.\vspace{-0.5cm}}}
    \label{fig:system_model}
\end{figure}
\par From an interference-centric viewpoint, \gls{isac} departs from conventional communication-only and sensing-only systems because several interference mechanisms coexist and interact \cite{Yuanwei_Uplink_ISaC1,Chiriyath2016,chen2025interference}. Recent literature broadly classifies these into clutter interference, self-interference, inter-target interference, inter-user interference, and \gls{ifi} between communication and sensing \cite{chen2025interference,Yuanwei_NOMA_ISaC,Wei_Survey}. Of these interference types, \gls{ifi} is particularly critical in highly integrated and non-orthogonal \gls{isac} architectures, where the two functionalities operate over shared resources and processing chains \cite{Chiriyath2016,Zhang2023SemiISAC}. In uplink bistatic \gls{isac} scenarios, as illustrated in Fig.~\ref{fig:system_model}, \gls{ifi} can become especially pronounced when the direct and echo paths contribute unequally to communication and sensing, with its severity jointly governed by their relative strengths and the receiver processing strategy \cite{Brunner2025,Schniter2004,Yu2025UplinkISACReceiver}. This imbalance-driven \gls{ifi} is further compounded in mobile \gls{ofdm} settings, where Doppler-induced loss of subcarrier orthogonality introduces an additional layer of interference in the form of \gls{ep} \gls{ici}\cite{Schniter2004,Zhang2020,Sahin2023}.
\par These considerations naturally motivate the use of multiple-access design principles for interference management in \gls{isac}. Taking inspiration from multiple access schemes managing \textit{inter-user interference} in communication-only systems, existing works have adapted strategies such as resource partitioning, and superposition and successive decoding to manage \gls{ifi} between coexisting sensing and communication signals \cite{Yuanwei_Uplink_ISaC1,Chiriyath2016,chen2025interference}. In this context, \gls{oma}-inspired \gls{isac} suppresses interference through resource separation, but at the cost of reduced \gls{se}. \Gls{noma}-inspired \gls{isac} improves resource utilization through superposition and \gls{sic}, yet inherits a rigid decoding structure that may leave one functionality strongly interference-limited by the other \cite{Yuanwei_Uplink_ISaC1,Yuanwei_NOMA_ISaC}. \am{To overcome these limitations, \gls{rs}-inspired \gls{isac} was introduced as a
more flexible framework that adapts the \gls{rs} principle to
inter-functionality interference management: part of the communication
message is decoded and cancelled before sensing, while the remaining part is
treated as interference during sensing and decoded subsequently
\cite{Mishra2025,Mishra2026RSInspiredUplink}.} Such flexibility is particularly attractive for managing \gls{ifi} in \gls{isac} settings where the sensing-communication coupling evolves across propagation conditions and receiver stages, while practical impairments further complicate the resulting interference structure, exposing the limitations of both \gls{oma}- and \gls{noma}-inspired strategies\cite{Mishra2025,Mishra@tutorial,Sahin2025,Mishra2022}.
\par Motivated by the above considerations, we propose and investigate an \gls{rs}-inspired uplink bistatic \gls{ofdm}-\gls{isac} framework in which a vehicular \gls{ue} transmits a superposed communication-sensing waveform to a network \gls{bs} while simultaneously illuminating a traffic target. The \gls{bs} acts as a dual-functional bistatic receiver, where the direct and target-echo paths jointly contribute to the received communication signal, while the echo path additionally conveys the target delay-Doppler information required for sensing. In mobile \gls{ofdm} settings, after \bc{compensation of the estimated bulk direct-path Doppler}, subcarrier orthogonality can be largely restored for the main communication link, whereas the residual bistatic Doppler of the echo remains uncompensated, causing Doppler-induced \gls{ici} across subcarriers \cite{Li_Yiheng,Brunner2025,Sahin2025}. Accordingly, the proposed framework addresses these coupled impairments by using \gls{rs}-inspired layered decoding, sensing, and interference cancellation to manage \gls{ifi}, while explicitly incorporating echo-path \gls{ici} into the sensing, reconstruction, and suppression stages. \bc{The two fixed-order \gls{noma}-inspired architectures are recovered as endpoint restrictions of the proposed formulation. Under a common channel-dependent sensing calibration, numerical results quantify the gain over their per-realization envelope, both at zero mean target Doppler and under target-induced echo-path \gls{ici}.}

\subsection{Related Works}
\label{sec:related_work}

Multiple-access-focused \gls{isac} research has developed along several complementary directions. On the one hand, {multiple-access-assisted} \gls{isac} schemes use multiple-access techniques primarily for inter-user interference management, while handling \gls{ifi} implicitly or through sensing--communication co-design. Reference \cite{Xu2021} proposes a multi-antenna \gls{rsma}-based dual-functional radar-communication architecture that improves the weighted-sum-rate and radar beampattern trade-off over \gls{sdma}-assisted and orthogonal baselines. Tutorial \cite{Longfei2022a} further reviews related extensions of this line, including \gls{rs}-based joint radar-communication transmission under partial \gls{csit} and \gls{rsma}-assisted dual-functional radar-communication satellite systems. Subsequently, \cite{Hu2023UplinkRSMAISAC} studies an uplink \gls{rsma}-enabled \gls{isac} system where \gls{rs} is applied across communication users to mitigate inter-user interference and enhance sensing performance over \gls{oma} and \gls{noma} baselines. On the other hand, multiple-access-inspired \gls{isac} works have addressed \gls{ifi} by explicitly borrowing multiple-access principles from communication-only systems. Reference \cite{Chiriyath2016} develops inner bounds for radar--communication coexistence in shared spectrum under isolated-subband and \gls{sic}-based operating regimes. Reference \cite{Wang2022NOMAInspired} studies a downlink \gls{isac} design in which part of the dedicated sensing signal is treated as virtual communication signals and mitigated via \gls{sic}, together with joint communication--sensing signal optimization. References \cite{Yuanwei_Uplink_ISaC1,Zhang2023SemiISAC} further develop downlink and uplink \gls{noma}-based \gls{isac} designs, including pure-\gls{noma}- and semi-\gls{noma}-based uplink architectures for mixed sensing--communication reception at the \gls{bs}. \am{More recently, \cite{Mishra2025} introduced an \gls{rs}-inspired coexistence framework, instantiated for delay estimation in a monostatic \gls{isac} setting, that applies the \gls{rs} principle by structuring the communication functionality around the sensing operation for flexible \gls{ifi} management. The uplink communication and sensing rate-region characterization of this \gls{rs}-inspired \gls{isac} framework, together with comparisons against \gls{oma}- and \gls{noma}-inspired schemes, is further examined in \cite{Mishra2026RSInspiredUplink}.} Pertinently, broader overviews of related multiple-access-focused \gls{isac} designs in various settings are provided in \cite{chen2025interference,Yuanwei_NOMA_ISaC,Longfei2022a}.
\par A separate but closely related line of work has focused on uplink bistatic and \gls{ofdm}-based \gls{isac}, together with receiver architectures relevant to this setting. In \cite{Li_Yiheng}, an uplink joint communication and sensing system with physically separated transceivers is analyzed, explicitly accounting for the impact of \gls{los}-path noise and communication decoding errors on delay--Doppler estimation. Bi-static sensing with \gls{5g} \gls{nr} physical uplink shared channel transmissions is investigated in \cite{Tapio2024}, which highlights practical bottlenecks such as limited \gls{ue} transmit power and the path-loss imbalance between the direct and echo links. A bistatic \gls{ofdm}-based \gls{isac} concept with over-the-air synchronization is then developed in \cite{Brunner2025}, where the impact of residual synchronization mismatches and communication decoding failures on both radar and communication performance is analyzed. Complementing these setting-focused works, \cite{Yu2025UplinkISACReceiver} develops a flexible uplink \gls{isac} receiver framework that bridges projection-type and \gls{sic}-type processing through a tunable tradeoff factor, highlighting the benefit of stage-adaptive receiver design. \bc{On the \gls{rs} side, \cite{Park2025bistatic} brings \gls{rsma} into a bistatic \gls{isac} setting for \gls{leo} satellite systems, where the common stream is jointly exploited for radar beamforming, inter-user interference management, and \gls{ifi} management under \gls{crlb} constraints. Finally, regarding robustness to Doppler-induced \gls{ici},
\cite{Sahin2025} demonstrates the benefits of \gls{rsma} in a
communication-only \gls{ofdm} setting, where message splitting enables
interference to be partially decoded and partially treated as
noise~\cite{Mishra@tutorial}. }
% ==========================================
\subsection{Motivation and Contributions}
% ==========================================
\bc{Despite recent progress on multiple-access-focused \gls{isac},
uplink bistatic sensing, and robust \gls{ofdm}-\gls{rsma} transmission,
these research directions have largely developed separately. In particular,
their intersection raises a distinct interference-management problem in
uplink bistatic \gls{ofdm}-\gls{isac}, where \gls{ifi} and echo-path
\gls{ici} must be jointly accounted for through receiver processing and
power allocation. This coupling is especially relevant because the direct
and reflected paths contribute unequally to communication and sensing,
while bistatic Doppler further complicates receiver operation
through \gls{ici}. Motivated by this setting and building upon the
\gls{rs}-inspired \gls{isac} framework developed
in~\cite{Mishra2025,Mishra2026RSInspiredUplink}, this paper makes the
following contributions.}
\begin{anup}
\begin{itemize}
    \item We develop an \gls{rs}-inspired uplink bistatic
    \gls{ofdm}-\gls{isac} framework in which a single communication
    message is split across the sensing stage into a robust stream,
    decoded before sensing, and a supplementary stream, decoded after
    sensing and echo reconstruction. Both streams are superposed with a
    dedicated sensing sequence under a common transmit-power budget.
    The two fixed-order \gls{noma}-inspired communication-first and
    sensing-first architectures arise as endpoint restrictions of this
    formulation, while the channel model captures the asymmetric
    \gls{dp}/\gls{ep} contributions and Doppler-induced echo-path
    \gls{ici}.

   \item We develop a staged receiver comprising \gls{dp}
sensing-sequence cancellation, robust-stream decoding, target-state
estimation, echo reconstruction, and supplementary-stream decoding.
Based on this structure, we derive per-subcarrier \gls{sinr}
expressions that account for predictable and unresolved echo components,
Doppler-induced \gls{ici}, sensing--communication interference, and
echo-reconstruction mismatch.

    \item We characterize the coupling between sensing uncertainty and post-sensing communication performance by explicitly accounting for the effect of target-state uncertainty on echo reconstruction. This captures how imperfect delay--Doppler knowledge translates into residual echo interference across the full \gls{ep} channel and, in turn, degrades supplementary-stream decoding. The resulting formulation therefore captures how sensing quality affects communication performance within the staged receiver.

    \item We formulate a sensing-constrained power-allocation problem that
jointly allocates power across the robust, supplementary, and sensing
components to balance communication performance with the required sensing
quality. To address the resulting non-convex design, we develop a tractable
iterative framework combining a \gls{lmi} reformulation of the sensing
constraint, a fixed-sensitivity reconstruction-mismatch approximation with
a local first-order correction, and multidimensional fractional programming,
together with full-model feasibility and objective auditing of candidate
updates.
    
   % \item Through numerical results, we show that the proposed \gls{rs}-inspired design outperforms \gls{noma}-inspired baselines across static and mobile regimes. In particular, it provides more flexible \gls{ifi} management in the zero-mobility case and exhibits greater robustness to \gls{ep} \gls{ici} under Doppler.
\end{itemize}
\end{anup}
% ==========================================
\par \textit{Notations}: 
Scalars are denoted by lowercase italic (\(x\)); vectors by bold lowercase (\(\mathbf{x}\)); parameter vectors by bold italic (e.g., \(\boldsymbol{\theta}\)); and matrices by bold uppercase (\(\mathbf{X}\)). The sets \(\mathbb{R}\), \(\mathbb{C}\), and \(\mathbb{Z}\) denote real numbers, complex numbers, and integers, respectively, \(\mathcal{CN}(\boldsymbol{\mu},\mathbf{\Sigma})\) denotes a \gls{cscg} distribution with mean \(\boldsymbol{\mu}\) and covariance \(\mathbf{\Sigma}\), and $\mathcal{U}[a, b)$ denotes the continuous uniform distribution over the interval $[a, b)$. We use \((\cdot)^*\) for complex conjugate, \((\cdot)^{\mathrm{H}}\) for Hermitian transpose, \(\mathbb{E}[\cdot]\) for expectation, and \(\operatorname{Re}\{\cdot\}\) for real part.
\par \textit{Organization}: The remainder of this paper is organized as follows. Section~\ref{sec:system_model} introduces the uplink bistatic \gls{ofdm}-\gls{isac} system model, including the \gls{rs}-inspired superposition transmission model. Section~\ref{sec:receiver_processing} presents the staged receiver, derives the per-subcarrier \gls{sinr} expressions, and develops the associated sensing-information surrogate and channel-reconstruction mismatch analyses. Section~\ref{sec:optimisation} formulates the sensing-constrained power-allocation problem. Section~\ref{sec:numerical_results} provides numerical results and benchmarks the proposed framework against \gls{noma}-inspired baselines. Finally, Section~\ref{sec:conclusion} concludes the paper.
% ==========================================
\section{System Model}
\label{sec:system_model}
% ==========================================
We consider the uplink bistatic \gls{v2n} \gls{isac} architecture
illustrated in Fig.~\ref{fig:system_model}. A vehicular \gls{ue}
transmits an uplink \gls{ofdm} waveform to a network \gls{bs} while
simultaneously illuminating a traffic target, e.g., another vehicle.
The \gls{bs} acts as a dual-functional bistatic receiver that decodes
the uplink communication streams and estimates the target delay and
Doppler from the reflected signal. Both the direct and echo paths carry
replicas of the transmitted superposed waveform. Owing to the bistatic
propagation loss and target scattering, however, the \gls{ep} is
typically much weaker than the \gls{dp} and additionally embeds the
target-dependent delay--Doppler information required for sensing
\cite{Willis2004,Tapio2024}. This path asymmetry creates a coupled
interference structure: strong direct-path communication components can
impair sensing before cancellation, while the target echo contributes
residual interference to communication decoding.
\begin{anup}
%==========================================
\subsection{\Gls{rs}-Inspired Superposition \gls{ofdm} Transmission}
%==========================================
Let $n\in\mathcal{N}\triangleq\{0,\ldots,N_{\textrm{sc}}-1\}$
denote the subcarrier index and $m\in\{0,\ldots,M-1\}$ the
\gls{ofdm} symbol index. We define the centered subcarrier coordinate
as $\kappa_n \triangleq n-(N_{\textrm{sc}}-1)/2$,
such that subcarrier $n$ has baseband frequency offset
$\kappa_n\Delta f$ relative to the band-center frequency $f_c$,
i.e., $f_n=f_c+\kappa_n\Delta f$. With the common band-center
phase included in the effective complex path coefficient, the
frequency-dependent delay phase is therefore determined by
$\kappa_n\Delta f$. For even $N_{\textrm{sc}}$, $\kappa_n$ is
half-integer and satisfies
$\sum_{n\in\mathcal{N}}\kappa_n=0$.

Following the \gls{rs}-inspired principle in
\cite{Mishra2025,Mishra2026RSInspiredUplink}, the communication
message is split into a robust stream $s_{c,1}$ and a supplementary
stream $s_{c,2}$. The robust stream is decoded before target-state
estimation, whereas the supplementary stream is decoded after sensing
and echo reconstruction. The two communication streams are
superposed with a sensing sequence, yielding the transmitted
frequency-domain symbol
\begin{equation}
\begin{split}
X_n[m]
=&
\sqrt{p_{c,1}[n]}\,s_{c,1}[n,m]
+
\sqrt{p_{c,2}[n]}\,s_{c,2}[n,m]
\\
&+
\sqrt{p_r[n]}\,s_r[n,m].
\end{split}
\label{eq:tx_symbol_n}
\end{equation}
Here, $s_{c,1}[n,m]$ and $s_{c,2}[n,m]$ denote the information
symbols of the robust and supplementary communication streams,
respectively, while $s_r[n,m]$ denotes the sensing symbol. The
communication symbols are mutually independent, independent of the
target state, and modeled as zero-mean and unit-power,
\begin{equation}
\mathbb{E}[s_{c,i}[n,m]]=0,
\quad
\mathbb{E}\!\left[|s_{c,i}[n,m]|^2\right]=1,
\quad i\in\{1,2\}.
\end{equation}
The sensing-sequence model is specified below; its realization is known
at the \gls{bs} during the considered block.

\textit{Assumption~1 (Sensing-sequence ensemble):}
The sensing symbols are modeled as $s_r[n,m]=e^{j\varphi_{n,m}},
\; \varphi_{n,m}\sim\mathcal{U}[0,2\pi)$,
with phases independent across subcarriers and symbols and independent
of the communication symbols and target state. The power-allocation
model uses the corresponding ensemble second-order statistics, whereas
a particular realization of the sensing sequence is fixed over the
considered block.\\[2ex]
The non-negative coefficients $p_{c,1}[n]$, $p_{c,2}[n]$, and
$p_r[n]$ denote the powers allocated to the robust stream,
supplementary stream, and sensing sequence, respectively, on subcarrier
$n$. They satisfy the per-\gls{ofdm}-symbol transmit-power constraint
\begin{equation}
\sum_{n\in\mathcal{N}}
\big(
p_{c,1}[n]+p_{c,2}[n]+p_r[n]
\big)
\le P_{\textrm{tx}},
\label{eq:power_constraint}
\end{equation}
where $P_{\textrm{tx}}$ denotes the maximum transmit power per
\gls{ofdm} symbol and
$P_{\textrm{avg}}\triangleq P_{\textrm{tx}}/N_{\textrm{sc}}$
the corresponding average available power per subcarrier. We collect
the allocation variables in
$\mathbf{p}_{c,1},\mathbf{p}_{c,2},\mathbf{p}_r
\in\mathbb{R}_{+}^{N_{\textrm{sc}}}$ and define
$\mathbf{p}\triangleq
(\mathbf{p}_{c,1},\mathbf{p}_{c,2},\mathbf{p}_r)$. Let
\begin{equation}
\mathbf{x}[m]
\triangleq
[X_0[m],\ldots,X_{N_{\textrm{sc}}-1}[m]]^T
\in\mathbb{C}^{N_{\textrm{sc}}\times1}
\end{equation}
denote the frequency-domain transmit vector, which we decompose as
$\mathbf{x}[m]
=
\mathbf{x}_1[m]
+
\mathbf{x}_2[m]
+
\mathbf{x}_r[m]$,
with entries
\begin{equation}
\begin{aligned}
[\mathbf{x}_1[m]]_n
&=
\sqrt{p_{c,1}[n]}\,s_{c,1}[n,m],\\
[\mathbf{x}_2[m]]_n
&=
\sqrt{p_{c,2}[n]}\,s_{c,2}[n,m],\\
[\mathbf{x}_r[m]]_n
&=
\sqrt{p_r[n]}\,s_r[n,m].
\end{aligned}
\end{equation}
Under the adopted symbol statistics, each component has a diagonal
second-moment matrix. In particular,
\begin{equation}
\mathbb{E}
\!\left[
\mathbf{x}_r[m]\mathbf{x}_r^H[m]
\right]
=
\operatorname{diag}
\big(
p_r[0],\ldots,p_r[N_{\textrm{sc}}-1]
\big).
\label{eq:sensing_symbol_covariance}
\end{equation}
Assuming that the \gls{cp} accommodates the delay support specified in
Section~\ref{subsec:received_signal}, the corresponding \gls{td}
transmit block is obtained through a unitary \gls{idft} followed by
\gls{cp} insertion,
\begin{equation}
\mathbf{x}_{\textrm{tx},\textrm{TD}}[m]
=
\mathbf{A}\mathbf{F}^H\mathbf{x}[m]
\in
\mathbb{C}^{(N_{\textrm{sc}}+N_{\textrm{cp}})\times1},
\label{eq:xtx_discrete_uplink}
\end{equation}
where
$\mathbf{F}\in
\mathbb{C}^{N_{\textrm{sc}}\times N_{\textrm{sc}}}$
is the unitary \gls{dft} matrix and
$\mathbf{A}\in
\mathbb{R}^{(N_{\textrm{sc}}+N_{\textrm{cp}})
\times N_{\textrm{sc}}}$
appends a \gls{cp} of length $N_{\textrm{cp}}$ \cite{Sahin2025}.
Specifically,
\begin{equation}
\mathbf{A}=
\begin{bmatrix}
\mathbf{0}_{N_{\textrm{cp}}\times
(N_{\textrm{sc}}-N_{\textrm{cp}})}
&
\mathbf{I}_{N_{\textrm{cp}}}
\\[2pt]
\multicolumn{2}{c}{\mathbf{I}_{N_{\textrm{sc}}}}
\end{bmatrix}.
\label{eq:A_matrix_def_uplink}
\end{equation}
Let
$L_{\textrm{blk}}
\triangleq
N_{\textrm{sc}}+N_{\textrm{cp}}$
denote the \gls{ofdm}-block length including the \gls{cp}.
The insertion and removal matrices satisfy
$\mathbf{A}\in
\mathbb{R}^{L_{\textrm{blk}}\times N_{\textrm{sc}}}$
and
$\mathbf{B}\in
\mathbb{R}^{N_{\textrm{sc}}\times L_{\textrm{blk}}}$,
while the cyclic delay operator $\mathbf{\Pi}$, Doppler operator
$\mathbf{\Delta}(\nu)$, and \gls{td} channel matrix
$\mathbf{H}_{\textrm{TD}}[m]$ belong to
$\mathbb{C}^{L_{\textrm{blk}}\times L_{\textrm{blk}}}$.
The corresponding effective frequency-domain channel is
\begin{equation}
\mathbf{H}_{\textrm{FD}}[m]
=
\mathbf{F}\mathbf{B}
\mathbf{H}_{\textrm{TD}}[m]
\mathbf{A}\mathbf{F}^H
\in
\mathbb{C}^{N_{\textrm{sc}}\times N_{\textrm{sc}}},
\label{eq:effective_fd_channel}
\end{equation}
whose diagonal entries represent intra-subcarrier responses and whose
off-diagonal entries capture \gls{ici}. When the \gls{cp}
accommodates the channel delay support and no residual Doppler is
present, $\mathbf{H}_{\textrm{FD}}[m]$ is diagonal. Finally, let
\begin{equation}
T_u\triangleq\frac{1}{\Delta f},
\quad
T_{\textrm{sym}}\triangleq T_u+T_{\textrm{CP}},
\quad
t_m\triangleq mT_{\textrm{sym}},
\end{equation}
denote the useful symbol duration, the complete \gls{ofdm} symbol
duration, and the reference time of symbol $m$, respectively. 

%==========================================
\subsection{\Gls{dp} Signal Model (\gls{ue} $\rightarrow$ \gls{bs})}
\label{subsec:dp_signal_model}
%==========================================
In discrete time, the \gls{dp} channel during the $m$-th \gls{ofdm}
symbol is represented as \cite{Sahin2025}
\begin{equation}
\mathbf{H}_{\textrm{TD}}^{(\mathrm{DP})}[m]
=
\sum_{l=1}^{L_{\textrm{p}}}
\alpha_l
\mathbf{\Pi}^{n_{\tau_l}}
\mathbf{\Delta}(\nu_l),
\label{eq:H_BS_td_matrix_def}
\end{equation}
where $L_{\textrm{p}}$ is the number of \gls{dp} taps, $\alpha_l$
denotes the complex gain of tap $l$, including small-scale fading,
propagation loss, and antenna gains, and $n_{\tau_l}$ is its discrete
delay index. The cyclic delay-shift matrix is denoted by
$\mathbf{\Pi}$, while the Doppler operator is given by
\begin{equation}
\mathbf{\Delta}(\nu_l)=\mathrm{diag}\!\left(
\big[
1,
e^{j\frac{2\pi\nu_l}{F_s}},
\ldots,
e^{j\frac{2\pi\nu_l(N_{\textrm{sc}}+N_{\textrm{cp}}-1)}{F_s}}
\big]^T
\right),
\label{eq:Delta_matrix_def_uplink}
\end{equation}
where $F_s\triangleq1/T_s$ is the sampling frequency
\cite{Sahin2025}. The corresponding noiseless \gls{td} received
signal is
\begin{equation}
\mathbf{y}_{\textrm{rx},\textrm{TD}}^{(\mathrm{DP})}[m]
=
\mathbf{H}_{\textrm{TD}}^{(\mathrm{DP})}[m]
\mathbf{x}_{\textrm{tx},\textrm{TD}}[m].
\label{eq:y_dp_td}
\end{equation}

After \gls{cp} removal by
$\mathbf{B}=
[\mathbf{0}_{N_{\textrm{sc}}\times N_{\textrm{cp}}}~~
\mathbf{I}_{N_{\textrm{sc}}}]$
and transformation to the \gls{fd}, the effective \gls{dp} channel is
\cite{Sahin2025}
\begin{equation}
\mathbf{H}_{\textrm{FD}}^{(\mathrm{DP})}[m]
\triangleq
\mathbf{F}\mathbf{B}
\mathbf{H}_{\textrm{TD}}^{(\mathrm{DP})}[m]
\mathbf{A}\mathbf{F}^H,
\label{eq:h_bs_direct_eff_definition}
\end{equation}
and the corresponding received component becomes
\begin{equation}
\mathbf{y}_{\textrm{rx},\textrm{FD}}^{(\mathrm{DP})}[m]
=
\mathbf{H}_{\textrm{FD}}^{(\mathrm{DP})}[m]\mathbf{x}[m].
\label{eq:y_dp_fd_vec}
\end{equation}
We denote the diagonal coefficient on subcarrier $n$ by
$H_{\textrm{DP},n}[m]
\triangleq
[\mathbf{H}_{\textrm{FD}}^{(\mathrm{DP})}[m]]_{n,n}$,
while the off-diagonal entries represent Doppler-induced \gls{ici}
\cite{Zhang2020,Schniter2004,Sahin2025}.

In general, the Doppler of tap $l$ can be written as
$\nu_l=\bar{\nu}_{\textrm{DP}}+\delta\nu_l$, where
$\bar{\nu}_{\textrm{DP}}$ denotes the bulk \gls{dp} Doppler removed
by receiver synchronization and $\delta\nu_l$ is the residual
tap-dependent Doppler. We consider the regime in which the bulk
\gls{dp} Doppler is compensated and the residual tap-dependent
Dopplers are negligible, and set $\delta\nu_l=0$ for all $l$ in the
numerical evaluation. This yields a diagonal but frequency-selective
\gls{dp} response when the \gls{cp} accommodates the delay spread,
while the target echo retains its residual bistatic Doppler and is
therefore modeled as non-diagonal in Section~\ref{subsec:echo_channel_model}.
This specialization is consistent with bistatic \gls{ofdm}-\gls{isac}
architectures synchronized to the dominant direct path
\cite{Brunner2025}.

Fractional tap delays are retained in the numerical frequency-domain
response and are not quantized to the sampling grid. Under the above
specialization,
\begin{equation}
H_{\textrm{DP},n}[m]
=
\sum_{l=1}^{L_{\textrm{p}}}
\alpha_l
\exp\!\left(
-j2\pi\kappa_n\Delta f\,\tau_l
\right).
\label{eq:dp_fractional_delay}
\end{equation}
Thus, the \gls{dp} retains frequency selectivity through the
continuous tap delays, while Doppler-induced \gls{ici} is confined to
the \gls{ep} in the adopted model. Non-negligible residual
tap-dependent \gls{dp} Doppler would require retaining the
off-diagonal terms of
$\mathbf{H}_{\textrm{FD}}^{(\mathrm{DP})}[m]$ and is outside the
present scope.
%==========================================
\subsection{Bistatic \gls{ep} Signal Model
(\gls{ue} $\rightarrow$ Target $\rightarrow$ \gls{bs})}
\label{subsec:echo_channel_model}
%==========================================

The \gls{bs} also receives a target-reflected bistatic echo. We model
the \gls{ep} by a single dominant bistatic path with effective complex
coefficient $\alpha_R\triangleq\bar{\alpha}_R g_R$, where
$\bar{\alpha}_R$ captures the large-scale gain and
$g_R=(\sqrt{K_R}+w_R)/\sqrt{K_R+1}$ is a Rician small-scale
coefficient, with $K_R$ denoting the linear Rician factor and
$w_R\sim\mathcal{CN}(0,1)$. Thus,
$\mathbb{E}[|g_R|^2]=1$, while the large-scale echo power follows the
bistatic radar equation
\begin{equation}
|\bar{\alpha}_R|^2
=
\frac{
G_{\textrm{UE}}G_{\textrm{BS}}
\lambda^2\sigma_{\textrm{RCS}}^{\textrm{bi}}
}{
(4\pi)^3R_{\textrm{UT}}^2R_{\textrm{TB}}^2
},
\label{eq:bistatic_reflection_coeff}
\end{equation}
where $G_{\textrm{UE}}$ and $G_{\textrm{BS}}$ denote the effective
antenna gains, $\lambda$ is the wavelength,
$\sigma_{\textrm{RCS}}^{\textrm{bi}}$ is the bistatic \gls{rcs}, and
$R_{\textrm{UT}}$ and $R_{\textrm{TB}}$ are the \gls{ue}-to-target
and target-to-\gls{bs} ranges, respectively \cite{Park2025bistatic}.
The corresponding bistatic propagation delay is
\begin{equation}
\tau_{\textrm{tar}}
=
\frac{R_{\textrm{UT}}+R_{\textrm{TB}}}{c},
\label{eq:bistatic_delay_target}
\end{equation}
where $c$ denotes the speed of light.

Since the bulk \gls{dp} Doppler is compensated at the receiver
\cite{Brunner2025}, the target echo is observed relative to this
reference with residual Doppler
\begin{equation}
\nu_{\textrm{tar}}
=
\nu_{\textrm{abs}}-\nu_{\textrm{DP}},
\label{eq:bistatic_doppler_relative}
\end{equation}
where $\nu_{\textrm{abs}}$ denotes the absolute bistatic Doppler and
$\nu_{\textrm{DP}}$ the compensated direct-path Doppler. For the
geometry in Fig.~\ref{fig:system_model}, the absolute bistatic Doppler
is \cite{Willis2004}
\begin{equation}
\nu_{\textrm{abs}}
=
\frac{\|\mathbf{v}_{\textrm{UE}}\|}{\lambda}
\cos(\delta_{\textrm{tx}})
+
\frac{2\|\mathbf{v}_{\textrm{tar}}\|}{\lambda}
\cos\phi\cos(\beta/2),
\label{eq:doppler_absolute}
\end{equation}
whereas the direct-path Doppler is
\begin{equation}
\nu_{\textrm{DP}}
=
\frac{\|\mathbf{v}_{\textrm{UE}}\|}{\lambda}
\cos(\delta_{\textrm{tx}}+\theta_T).
\label{eq:doppler_dp}
\end{equation}
Here, $\mathbf{v}_{\textrm{UE}}$ and $\mathbf{v}_{\textrm{tar}}$
denote the \gls{ue} and target velocities, respectively;
$\delta_{\textrm{tx}}$ and $\delta_{\textrm{tx}}+\theta_T$ are the
corresponding \gls{ue} velocity-projection angles for the echo and
direct paths, $\beta$ is the bistatic angle, and $\phi$ is measured
from its internal bisector to the target velocity vector. Positive
Doppler corresponds to decreasing propagation distance. We consider
target geometries for which the echo excess delay relative to the
direct-path timing reference is accommodated by the \gls{cp}; the
corresponding prior-based condition is stated in
Section~\ref{subsec:received_signal}. 

The coefficient $\alpha_R$ is an effective complex baseband echo
coefficient referenced to the band center and the first \gls{ofdm}
symbol, while \eqref{eq:bistatic_reflection_coeff} specifies its
average power.

\textit{Assumption~2 (Known effective echo coefficient):}
A preceding acquisition stage provides the effective complex echo
coefficient $\alpha_R$, which is treated as known and constant over the
considered processing interval. The current target state
$\boldsymbol{\theta}\triangleq
[\tau_{\textrm{tar}},\nu_{\textrm{tar}}]^T$
remains unknown and is described by the prior model in
Section~\ref{subsec:prior_information}. Such a staged abstraction is
consistent with task-conditioned sensing, where selected target or
channel information is acquired prior to the current sensing operation
\cite{Chiriyath2016,Zhang2023SemiISAC}.
This allows the present analysis to focus on \gls{ifi} and
Doppler-induced \gls{ici} management across the considered
non-orthogonal processing strategies. Accordingly, $\alpha_R$ is held
fixed when averaging over or differentiating with respect to
$\boldsymbol{\theta}$.

Under this model, the effective \gls{fd} echo channel is written as
$\mathbf{H}_{\textrm{FD}}^{(\mathrm{EP})}[m]
=\alpha_R\mathbf{S}[m;\boldsymbol{\theta}]$, whose $(n,k)$-th entry is
\begin{equation}
S_{n,k}[m;\boldsymbol{\theta}]
=
e^{j2\pi t_m\nu_{\textrm{tar}}}
c_{n,k}(\varepsilon_{\textrm{tar}})
e^{-j2\pi\kappa_k\Delta f\,\tau_{\textrm{tar}}},
\label{eq:h_tar_eff_definition}
\end{equation}
where
$\varepsilon_{\textrm{tar}}
\triangleq\nu_{\textrm{tar}}/\Delta f$.
The Doppler-induced coupling coefficient from source subcarrier $k$ to
observation subcarrier $n$ is \cite{Moose1994,Wang2023,Brunner2025}
\begin{equation}
c_{n,k}(\varepsilon_{\textrm{tar}})
=
\frac{1}{N_{\textrm{sc}}}
\sum_{q=0}^{N_{\textrm{sc}}-1}
e^{j2\pi(k-n)q/N_{\textrm{sc}}}
e^{j2\pi\varepsilon_{\textrm{tar}}q/N_{\textrm{sc}}}.
\label{eq:kernel_dft_representation}
\end{equation}
The coefficient $c_{n,k}$ captures Doppler-induced coupling over the
useful interval $T_u$, while $t_m=mT_{\textrm{sym}}$ accounts for the
full \gls{ofdm} symbol spacing. The delay-dependent phase is indexed by
the source subcarrier $k$.

The corresponding echo contribution to the received \gls{fd} vector is
\begin{equation}
\mathbf{y}_{\textrm{rx},\textrm{FD}}^{(\mathrm{EP})}[m]
=
\mathbf{H}_{\textrm{FD}}^{(\mathrm{EP})}[m]\mathbf{x}[m].
\label{eq:y_ep_fd_vec}
\end{equation}
On subcarrier $n$, this becomes
\begin{equation}
y_{\textrm{rx},\textrm{FD},n}^{(\mathrm{EP})}[m]
=
\underbrace{
H_{\textrm{FD},n,n}^{(\mathrm{EP})}[m]X_n[m]
}_{\text{Target return}}
+\!\!\!\!\!\!
\underbrace{
\sum_{k\in\mathcal{N},\,k\neq n}
H_{\textrm{FD},n,k}^{(\mathrm{EP})}[m]X_k[m]
}_{\text{Doppler-induced echo ICI}}.
\label{eq:rx_echo_subcarrier_n_ICI}
\end{equation}
The diagonal term represents the delay--Doppler-dependent
same-subcarrier response, including the diagonal Doppler attenuation,
whereas the off-diagonal terms capture Doppler-induced leakage across
subcarriers. We denote the diagonal echo coefficient by
$H_{\textrm{EP},n}[m]\triangleq
[\mathbf{H}_{\textrm{FD}}^{(\mathrm{EP})}[m]]_{n,n}$.
Since
$|S_{n,k}[m;\boldsymbol{\theta}]|
=
|c_{n,k}(\varepsilon_{\textrm{tar}})|$,
the redistribution of echo power across subcarriers is governed by the
residual Doppler, whereas the continuous delay contributes a
subcarrier-dependent phase rotation and does not itself generate
\gls{ici}.

%==========================================
\subsection{Prior Target-State Information and Echo Statistics}
\label{subsec:prior_information}
%==========================================

The target state available before transmission is statistical rather
than realization-specific. This setting follows prior-informed sensing
models in which the parameter to be estimated is treated as an unknown
random quantity whose distribution is available for waveform or
resource design \cite{XuZhang2024PriorISAC}. Such prior information may,
for instance, originate from a preceding acquisition stage,
a target-motion model, or historical observations. For the current
block, we model the target delay and residual Doppler as independent
Gaussian quantities,
\begin{equation}
\tau_{\textrm{tar}}
\sim
\mathcal{N}(\bar{\tau},\sigma_\tau^2),
\qquad
\nu_{\textrm{tar}}
\sim
\mathcal{N}(\bar{\nu},\sigma_\nu^2),
\label{eq:prior_model}
\end{equation}
where $\bar{\tau}$ and $\bar{\nu}$ denote the delay and Doppler
associated with the acquisition reference, while $\sigma_\tau$ and
$\sigma_\nu$ quantify the corresponding uncertainty. The Gaussian
specialization provides a local probabilistic description around this
reference and, importantly, permits the echo-channel moments required
for allocation to be obtained in closed form. The realized
$\boldsymbol{\theta}
=[\tau_{\textrm{tar}},\nu_{\textrm{tar}}]^T$
still determines the physical echo within the block. Hence,
$\bar{\nu}=0$ denotes zero prior-mean residual Doppler and does not
imply a static target.

Substituting \eqref{eq:kernel_dft_representation} into
\eqref{eq:h_tar_eff_definition} expresses each echo-channel entry as a
finite sum of exponentials linear in $\tau_{\textrm{tar}}$ and
$\nu_{\textrm{tar}}$. Defining
$a_k\triangleq-\kappa_k\Delta f$ and
$b_{m,q}\triangleq t_m+q/(N_{\textrm{sc}}\Delta f)$, the Gaussian
characteristic function gives the conditional mean
\[
\overline{H}_{\textrm{EP},n,k}[m]
\triangleq
\mathbb{E}\!\left[
[\mathbf{H}_{\textrm{FD}}^{(\mathrm{EP})}[m]]_{n,k}
\mid\mathcal{I}^{-}
\right]
\]
as
\begin{equation}
\begin{aligned}
\overline{H}_{\textrm{EP},n,k}[m]
&=
\frac{\alpha_R}{N_{\textrm{sc}}}
\sum_{q=0}^{N_{\textrm{sc}}-1}
e^{j2\pi(k-n)q/N_{\textrm{sc}}}
e^{j2\pi(a_k\bar{\tau}+b_{m,q}\bar{\nu})}
\\
&\qquad\qquad\qquad\qquad\times
e^{-2\pi^2
\left(
a_k^2\sigma_\tau^2+
b_{m,q}^2\sigma_\nu^2
\right)} .
\end{aligned}
\label{eq:prior_mean_echo}
\end{equation}
Thus, the prior generally provides a nonzero predictable echo
component before transmission, rather than reducing the entire echo to
zero-mean interference. The corresponding prior-averaged transfer power from source subcarrier $k$ to observation subcarrier $n$ is
\begin{equation}
L_{n,k}
\triangleq
\mathbb{E}\!\left[
\left|
[\mathbf{H}_{\textrm{FD}}^{(\mathrm{EP})}[m]]_{n,k}
\right|^2
\mid\mathcal{I}^{-}
\right]
=
|\alpha_R|^2\Lambda(k-n),
\label{eq:prior_leakage_definition}
\end{equation}
where, with
$\mu_\varepsilon\triangleq\bar{\nu}/\Delta f$ and
$\sigma_\varepsilon\triangleq\sigma_\nu/\Delta f$,
\begin{equation}
\begin{aligned}
\Lambda(d)
&=
\frac{1}{N_{\textrm{sc}}}
+
\frac{2}{N_{\textrm{sc}}^2}
\sum_{s=1}^{N_{\textrm{sc}}-1}
(N_{\textrm{sc}}-s)
\\
&\qquad \qquad\times
\cos\!\left(
\frac{2\pi s(d+\mu_\varepsilon)}
{N_{\textrm{sc}}}
\right)
e^{-2\pi^2s^2\sigma_\varepsilon^2/
N_{\textrm{sc}}^2}.
\end{aligned}
\label{eq:lambda_kernel}
\end{equation}
The matrix $\mathbf{L}\triangleq[L_{n,k}]$ is circulant, with every
row and column summing to $|\alpha_R|^2$. The target-Doppler prior
therefore redistributes echo power among the subcarriers while
preserving the total echo power.

The prior mean in \eqref{eq:prior_mean_echo} also provides a natural
decomposition of the echo channel into predictable and unresolved
components,
$[\mathbf{H}_{\textrm{FD}}^{(\mathrm{EP})}[m]]_{n,k}
=
\overline{H}_{\textrm{EP},n,k}[m]
+
\widetilde{H}_{\textrm{EP},n,k}[m]$.
This known-component-plus-uncertainty representation follows the
statistical treatment commonly used for imperfect channel knowledge
\cite{Medard2000}; here, however, the uncertainty originates from the
unknown target state rather than from communication-channel estimation
error. Conditioned on $\mathcal{I}^{-}$,
$\widetilde{H}_{\textrm{EP},n,k}[m]$ has zero mean and variance
\begin{equation}
\begin{split}
V_{n,k}[m]
\triangleq &
\mathbb{E}\!\left[
|\widetilde{H}_{\textrm{EP},n,k}[m]|^2
\mid\mathcal{I}^{-}
\right] \\
=&
L_{n,k}
-
|\overline{H}_{\textrm{EP},n,k}[m]|^2.
\end{split}
\label{eq:echo_prior_variance}
\end{equation}
No independence among the echo-channel entries is required by this
decomposition. Together with the symbol statistics in Assumption~1,
the second moments yield prior-averaged echo powers that are affine in
the allocation, e.g.,
$\sum_{k\in\mathcal{N}}L_{n,k}p_r[k]$ for the sensing component.

The same prior distribution also contributes information about the
unknown target state before observing the current block. Consistent
with Bayesian sensing formulations that combine prior and observation
information \cite{XuZhang2024PriorISAC}, the independent Gaussian
prior in \eqref{eq:prior_model} yields
$\mathbf{J}_p
=
\operatorname{diag}
(\sigma_\tau^{-2},\sigma_\nu^{-2})$.
This matrix regularizes the observation-based sensing-information
surrogate introduced in Section~\ref{subsec:CRLB}. To place delay and
Doppler on a common dimensionless scale, we define
$\mathbf{W}
\triangleq
\operatorname{diag}
(\tau_{\textrm{ref}}^{-2},\nu_{\textrm{ref}}^{-2})$,
where $\tau_{\textrm{ref}}$ and $\nu_{\textrm{ref}}$ are fixed
reference scales common to all architectures. The resulting prior-only
sensing score is
\begin{equation}
\Gamma_{\textrm{prior}}
\triangleq
\operatorname{Tr}
\left(
\mathbf{W}\mathbf{J}_p^{-1}
\right)
=
\frac{\sigma_\tau^2}{\tau_{\textrm{ref}}^2}
+
\frac{\sigma_\nu^2}{\nu_{\textrm{ref}}^2},
\label{eq:prior_floor}
\end{equation}
which is the sensing score available from the prior alone, before
information from the current observation is incorporated.

%==========================================
\subsection{Received Signal and Pre-Transmission Information}
\label{subsec:received_signal}
%==========================================

We assume symbol-level synchronization at the \gls{bs} with respect
to the \gls{dp}, with the \gls{cp} accommodating its delay spread,
consistent with closed-loop timing advance in uplink systems
\cite{Chen2020,Tapio2024}. The received \gls{fd} signal on subcarrier
$n$ is
\begin{equation}
y_{\textrm{rx},\textrm{FD},n}[m]
=
y_{\textrm{rx},\textrm{FD},n}^{(\mathrm{DP})}[m]
+
y_{\textrm{rx},\textrm{FD},n}^{(\mathrm{EP})}[m]
+
z_{\textrm{BS},\textrm{FD},n}[m],
\label{eq:total_received_signal}
\end{equation}
where
$z_{\textrm{BS},\textrm{FD},n}[m]
\sim\mathcal{CN}(0,\sigma_{\textrm{BS}}^2)$
denotes \gls{awgn} with variance $\sigma_{\textrm{BS}}^2$.

For the random target delay in \eqref{eq:prior_model}, we restrict the
considered operating regime to
$\bar{\tau}+3\sigma_\tau\le\tau_{\textrm{CP}}$, where
$\tau_{\textrm{CP}}
\triangleq d_{\textrm{UE-BS}}/c+T_{\textrm{CP}}$.
Under the Gaussian prior, this corresponds to
$\Pr\{\tau_{\textrm{tar}}>\tau_{\textrm{CP}}\}
\le Q(3)\approx1.35\times10^{-3}$.
Target realizations outside this coverage are therefore outside the
adopted \gls{cp}-based channel model.

The effective diagonal \gls{dp} coefficients are acquired from known
uplink pilots once per coherence block. As specified in
Section~\ref{subsec:dp_signal_model}, the bulk \gls{dp} Doppler is
compensated, while the target echo retains the residual Doppler
$\nu_{\textrm{tar}}$ in \eqref{eq:bistatic_doppler_relative}.
The resulting \gls{dp} \gls{csi} is used during receiver processing to
remove the direct-path replicas of known signal components, consistent
with direct-path suppression in uplink bistatic sensing and passive
radar \cite{Li_Yiheng,Willis2004,Garry2017}. We assume perfect effective
\gls{dp} \gls{csi},
$\widehat{H}_{\textrm{DP},n}[m]
=
H_{\textrm{DP},n}[m]$,
and defer the effect of imperfect cancellation to Remark~1.

Before transmission of block $t$, the information available for
power allocation is
\begin{equation}
\mathcal{I}^{-}
=
\big\{
\widehat{\mathbf{H}}_{\textrm{DP},t},
\alpha_R,
\bar{\tau},
\sigma_\tau,
\bar{\nu},
\sigma_\nu
\big\},
\quad
\mathbf{p}_t^{\star}=f(\mathcal{I}^{-}),
\label{eq:information_set}
\end{equation}
where $\widehat{\mathbf{H}}_{\textrm{DP},t}$ collects the effective
\gls{dp} coefficients for block $t$, and
$\mathbf{p}_t^{\star}$ collects the robust-stream, supplementary-stream,
and sensing power allocations. All echo-dependent design quantities are
computed from $\alpha_R$ and the prior statistics in
Section~\ref{subsec:prior_information}; the realized
$\tau_{\textrm{tar}}$ and $\nu_{\textrm{tar}}$, and their current-block
estimates, are therefore not used in the same-block allocation.

The resulting power allocations are signaled to the \gls{ue} before
data transmission. Quantization, control signaling errors, and
associated latency are outside the present model. We adopt a
\gls{siso}-equivalent link under beam alignment, with array gains
absorbed into the link budget.

\begin{figure*}[t!]
\centering
\includegraphics[width=\textwidth]{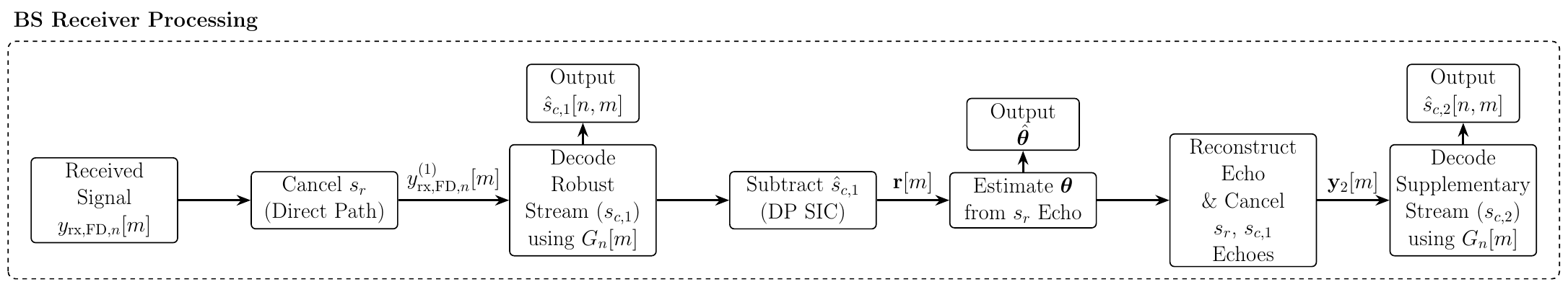}
\caption{\bc{Proposed \gls{rs}-inspired receiver at the \gls{bs}: \gls{dp} sensing-replica cancellation, robust-stream decoding and removal of its \gls{dp} contribution, sensing-sequence-based target-state estimation, cancellation of the reconstructed known-component echoes, and supplementary-stream decoding. Both streams use the prior-statistical coefficient $G_n[m]$ in \eqref{eq:effective_statistical_channel}.}}
\vspace{-0.5cm}
\label{fig:processing_flow}
\end{figure*}

% ==========================================
\section{Receiver Processing and Effective \gls{sinr} Analysis}
\label{sec:receiver_processing}
% ==========================================

The staged receiver in Fig.~\ref{fig:processing_flow} follows the
\gls{rs}-inspired processing order of \cite{Mishra2025}. The
\gls{bs} first removes the known \gls{dp} sensing replica and decodes
the robust communication stream. After cancelling the robust stream's
\gls{dp} contribution, the target state is estimated from the sensing
reference and used to reconstruct and suppress the echoes of the known
signal components. The supplementary communication stream is then
decoded from the resulting observation. Both communication stages use
the prior-statistical echo information developed in
Section~\ref{subsec:prior_information}; the current-block target-state
estimate is used only for echo reconstruction and cancellation.

The same statistical channel model and sensing criterion are used for
all considered architectures. Setting
$\mathbf{p}_{c,2}=\mathbf{0}$ recovers \gls{noma}-CF, for which the
entire communication message is decoded before sensing, whereas
$\mathbf{p}_{c,1}=\mathbf{0}$ recovers \gls{noma}-SF, for which the
entire message is decoded after sensing and echo recovery. The
\gls{rs}-inspired design permits both streams to be active and
therefore interpolates between these two fixed processing orders.
The blockwise \gls{sinr} quantities introduced below serve as effective
metrics for power allocation.

% =============================================================================
\paragraph{Step 1: \gls{dp} Sensing-Sequence Cancellation}
% =============================================================================

Using the known sensing sequence $s_r[n,m]$ and the \gls{dp}
coefficient $H_{\textrm{DP},n}[m]$, the \gls{bs} first removes the
\gls{dp} replica of the sensing component \cite{Sturm2011}. Under
perfect \gls{dp} cancellation, the resulting signal on subcarrier $n$
is
\begin{align}
y_{\textrm{rx},\textrm{FD},n}^{(1)}[m]
&=
y_{\textrm{rx},\textrm{FD},n}[m]
-
H_{\textrm{DP},n}[m]\sqrt{p_r[n]}\,s_r[n,m]
\nonumber\\
&=
\tilde{y}_{\textrm{rx},\textrm{FD},n}^{(\mathrm{DP,c})}[m]
+
y_{\textrm{rx},\textrm{FD},n}^{(\mathrm{EP})}[m]
+
z_{\textrm{BS},\textrm{FD},n}[m],
\label{eq:rx_after_dp_radar_cancel}
\end{align}
where
$\tilde{y}_{\textrm{rx},\textrm{FD},n}^{(\mathrm{DP,c})}[m]$
contains the remaining \gls{dp} communication components,
\begin{equation}
\begin{split}
\tilde{y}_{\textrm{rx},\textrm{FD},n}^{(\mathrm{DP,c})}[m]
=&
H_{\textrm{DP},n}[m]
\Big(
\sqrt{p_{c,1}[n]}s_{c,1}[n,m]
\\
&\hspace{1.6cm}
+
\sqrt{p_{c,2}[n]}s_{c,2}[n,m]
\Big).
\label{eq:dp_comms_residual}
\end{split}
\end{equation}
Only the \gls{dp} sensing replica is cancelled at this stage. Its
target-reflected counterpart remains because it propagates through the
distinct bistatic channel
$\mathbf{H}_{\textrm{FD}}^{(\mathrm{EP})}[m]$
\cite{Garry2017,Willis2004}.

\textit{Remark~1 (Imperfect direct-path cancellation):}
The numerical evaluation assumes perfect effective \gls{dp} \gls{csi},
i.e.,
$\widehat{H}_{\textrm{DP},n}[m]=H_{\textrm{DP},n}[m]$,
and omits pilot overhead. If
$\widehat{H}_{\textrm{DP},n}[m]
=
H_{\textrm{DP},n}[m]+\Delta H_{\textrm{DP},n}[m]$,
the sensing-sequence cancellation leaves the residual
$\Delta y_n[m]
=
-\Delta H_{\textrm{DP},n}[m]
\sqrt{p_r[n]}s_r[n,m]$.
For zero-mean estimation error with variance
$\sigma_{\Delta,n}^2$, this contributes
$\sigma_{\Delta,n}^2p_r[n]$ during robust-stream decoding and
$\sigma_{\Delta,n}^2(p_r[n]+p_{c,1}[n])$ to the sensing observation
after robust-stream cancellation. Such residual \gls{dp} interference
is outside the present comparison.

%=============================================================
\paragraph{Step 2: Decoding of the Robust Stream ($s_{c,1}$)}
%=============================================================
The \gls{bs} decodes $s_{c,1}[n,m]$ from
$y_{\textrm{rx},\textrm{FD},n}^{(1)}[m]$, while the supplementary
stream remains undecoded and is treated as interference.

\textit{Assumption~3 (Prior-statistical communication detection):}
Both communication streams are detected using the same
prior-statistical diagonal channel coefficient
\begin{equation}
G_n[m]
\triangleq
H_{\textrm{DP},n}[m]
+
\overline{H}_{\textrm{EP},n,n}[m].
\label{eq:effective_statistical_channel}
\end{equation}
Thus, the statistically predictable diagonal echo component is included
coherently in communication detection, while the zero-mean unresolved
component and the off-diagonal echo terms are treated as disturbance.
This known-component-plus-residual representation is consistent with
the standard treatment of imperfect channel knowledge in
\cite{Medard2000}. The current-block reconstruction
$\widehat{\mathbf{H}}_{\textrm{FD}}^{(\mathrm{EP})}[m]$
is used only for cancellation and not as coherent communication
\gls{csi}. For blockwise allocation, define
\begin{equation}
\begin{aligned}
\mathcal{G}_n
&\triangleq
\frac{1}{M}\sum_{m=0}^{M-1}\left|G_n[m]\right|^2,\\
\mathcal{V}_n
&\triangleq
\frac{1}{M}\sum_{m=0}^{M-1}V_{n,n}[m],\\
\mathcal{Q}_n
&\triangleq
\mathcal{G}_n+\mathcal{V}_n.
\end{aligned}
\label{eq:block_statistical_coefficients}
\end{equation}
where $\mathcal{G}_n$ is the block-averaged coherent-channel power,
$\mathcal{V}_n$ is the unresolved diagonal echo variance, and
$\mathcal{Q}_n$ is the corresponding total prior-average
same-subcarrier power. All three quantities are determined before
transmission from $\mathcal{I}^{-}$. The useful robust-stream power is
\begin{equation}
S_{c,1,n}
\triangleq
\mathcal{G}_n p_{c,1}[n].
\label{eq:Sc1n}
\end{equation}
The unresolved component of the robust-stream echo contributes
$\mathcal{V}_np_{c,1}[n]$, whereas the undecoded supplementary stream
contributes its complete same-subcarrier power
$\mathcal{Q}_np_{c,2}[n]$. Accounting additionally for off-diagonal
communication echoes and the uncancelled sensing echo gives
\begin{equation}
\begin{aligned}
I_{c,1,n}
\triangleq &
\mathcal{V}_np_{c,1}[n]
+
\mathcal{Q}_np_{c,2}[n]
+
\!\!\!\!\!\!\! \\
&\sum_{k\in\mathcal{N}\setminus\{n\}}
\!\!\!\!\!\!\!
L_{n,k}
\big(
p_{c,1}[k]+p_{c,2}[k]
\big)
+
\sum_{k\in\mathcal{N}}
L_{n,k}p_r[k].
\end{aligned}
\label{eq:Ic1n_diagonal}
\end{equation}
The effective robust-stream \gls{sinr} is therefore
\begin{equation}
\gamma_{c,1,n}
\triangleq
\frac{S_{c,1,n}}
{I_{c,1,n}+\sigma_{\textrm{BS}}^2}.
\label{eq:gamma_c1_n}
\end{equation}
The coherent term $\mathcal{G}_n$ retains the relative phase between
the \gls{dp} and the prior echo mean; no phase alignment between them is
assumed. Hence, the prior echo mean may either increase or decrease the
coherent same-subcarrier power. Equation~\eqref{eq:gamma_c1_n} is based
on block-averaged powers and is used as an allocation metric rather
than an average of symbolwise rates.

After successful decoding, the \gls{bs} regenerates
$\hat{s}_{c,1}[n,m]$ and cancels its \gls{dp} contribution. Its
target-reflected component remains in the observation used for sensing.

%=============================================================================
\paragraph{Step 3: Target-State Estimation}
%=============================================================================

After removal of the robust stream's \gls{dp} component, and assuming
correct decoding and perfect effective \gls{dp} \gls{csi}, the
frequency-domain observation available for sensing is
\begin{equation}
\begin{aligned}
\mathbf{r}[m]
&=
\alpha_R\mathbf{S}[m;\boldsymbol{\theta}]
\big(
\mathbf{x}_r[m]
+
\mathbf{x}_1[m]
+
\mathbf{x}_2[m]
\big)
\\
&\quad+
\mathbf{H}_{\textrm{FD}}^{(\mathrm{DP})}[m]
\mathbf{x}_2[m]
+
\mathbf{z}[m],
\end{aligned}
\label{eq:sensing_observation}
\end{equation}
where $\mathbf{z}[m]$ collects the receiver-noise samples. At this
stage, $\alpha_R$,
$\mathbf{H}_{\textrm{FD}}^{(\mathrm{DP})}[m]$,
$\mathbf{x}_r[m]$, and the decoded $\mathbf{x}_1[m]$ are known,
whereas $\boldsymbol{\theta}$ and $\mathbf{x}_2[m]$ remain unknown.

\textit{Assumption~4 (Dedicated sensing reference):}
Target-state estimation is modeled using the sensing sequence
$\mathbf{x}_r[m]$ as the probing reference. The echo of the decoded
robust stream is therefore not exploited as a sensing reference and is
included in the disturbance budget.

Given an estimate $\widehat{\boldsymbol{\theta}}$, the receiver forms
$\widehat{\mathbf{H}}_{\textrm{FD}}^{(\mathrm{EP})}[m]
=
\alpha_R
\mathbf{S}[m;\widehat{\boldsymbol{\theta}}]$
for the subsequent echo-cancellation stage. For power allocation,
sensing quality is represented by the prior-regularized surrogate
developed in Section~\ref{subsec:CRLB}. Its useful per-subcarrier term
is
\begin{equation}
S_{r,n}
\triangleq
L_{n,n}p_r[n],
\label{eq:Sr_n}
\end{equation}
where $L_{n,n}=|\alpha_R|^2\Lambda(0)$. The undecoded supplementary
stream contributes through its complete \gls{dp} and \gls{ep} channel,
whereas the robust stream retains only its echo contribution after
\gls{dp} cancellation. Including sensing self-\gls{ici}, the
corresponding disturbance term is
\begin{equation}
\begin{aligned}
I_{r,n}
&\triangleq
\mathcal{Q}_np_{c,2}[n]
+
 \\
&
\sum_{k\in\mathcal{N}\setminus\{n\}}
\!\!L_{n,k}
\big(
p_{c,2}[k]+p_r[k]
\big)
+
\sum_{k\in\mathcal{N}}
L_{n,k}p_{c,1}[k].
\end{aligned}
\label{eq:Irn_diagonal}
\end{equation}
The resulting sensing-design weight is
\begin{equation}
\gamma_{r,n}
\triangleq
\frac{S_{r,n}}
{I_{r,n}+\sigma_{\textrm{BS}}^2}.
\label{eq:gamma_r_n}
\end{equation}

\textit{Remark 2 (Scope of the sensing weight):}
The quantity $\gamma_{r,n}$ is a design weight used to construct the
sensing-information surrogate, rather than the output SINR of a
particular estimator. Under Assumption~4, the decoded robust-stream echo
and sensing self-\gls{ici} are not exploited as additional target-state
information. Demodulate-then-sense processing can instead reuse decoded
communication symbols as sensing references~\cite{Dai2026}; however,
under the retained non-diagonal \gls{ep} channel, doing so would require
an \gls{ici}-aware composite-reference sensing model and a corresponding
reformulation of the power-allocation problem. Consequently,
$\mathbf{p}_r=\mathbf{0}$ removes the observation contribution in the
adopted sensing surrogate.

% =============================================================================
\paragraph{Step 4: Decoding of the Supplementary Stream ($s_{c,2}$)}
% =============================================================================

Using $\widehat{\boldsymbol{\theta}}$ obtained in Step~3, the \gls{bs}
reconstructs
$\widehat{\mathbf{H}}_{\textrm{FD}}^{(\mathrm{EP})}[m]$
and cancels the echo replicas of the known components
$\mathbf{u}[m]\triangleq\mathbf{x}_r[m]+\mathbf{x}_1[m]$.
The reconstructed echo channel is used only for this cancellation;
supplementary-stream detection continues to use the
coefficient $G_n[m]$ in
\eqref{eq:effective_statistical_channel}. Its useful power is
\begin{equation}
S_{c,2,n}
\triangleq
\mathcal{G}_np_{c,2}[n].
\label{eq:Sc2n}
\end{equation}
Imperfect echo reconstruction leaves a residual associated with the
cancelled components. Its power is represented by the local mismatch
model developed in Section~\ref{subsec:channel_error_analysis},
\begin{equation}
P_{\textrm{mismatch},n}(\mathbf{p})
\triangleq
\sum_{k\in\mathcal{N}}
\widetilde{\Omega}_k(\mathbf{p})
L_{n,k}
\big(
p_{c,1}[k]+p_r[k]
\big),
\label{eq:mismatch_power}
\end{equation}
where $\widetilde{\Omega}_k(\mathbf{p})$ is the block-averaged
reconstruction sensitivity defined in
\eqref{eq:omega_block_average}. Only the powers of the known,
cancelled components enter this mismatch term.

The supplementary-stream echo itself remains uncancelled. Since its
prior diagonal mean is already included in $G_n[m]$, its disturbance
contribution consists of the unresolved diagonal variance,
off-diagonal \gls{ici}, and the reconstruction mismatch. Accordingly,
\begin{equation}
I_{c,2,n}(\mathbf{p})
\triangleq
\mathcal{V}_np_{c,2}[n]
+
\!\!\!\!\!\!\!
\sum_{k\in\mathcal{N}\setminus\{n\}}
\!\!\!\!\!\!\!
L_{n,k}p_{c,2}[k]
+
P_{\textrm{mismatch},n}(\mathbf{p}),
\label{eq:Ic2_statistical}
\end{equation}
and the effective supplementary-stream \gls{sinr} is
\begin{equation}
\gamma_{c,2,n}
\triangleq
\frac{S_{c,2,n}}
{I_{c,2,n}(\mathbf{p})+\sigma_{\textrm{BS}}^2}.
\label{eq:sinr_c2_final}
\end{equation}
The sensing and supplementary-decoding stages are therefore coupled
through $P_{\textrm{mismatch},n}(\mathbf{p})$: the communication
allocation affects the sensing disturbance, which determines the
reconstruction sensitivity and hence the residual interference seen by
the supplementary stream. The construction and scope of this local
mismatch model are detailed in
Section~\ref{subsec:channel_error_analysis}.
% ==========================================
\subsection{Prior-Regularized Sensing Information Surrogate}
\label{subsec:CRLB}
% ==========================================

For tractable power allocation, we construct the sensing-information
surrogate from the explicit delay--Doppler phase of the retained
diagonal echo component,
\begin{equation}
\psi_{n,m}
=
2\pi
\big(
-\kappa_n\Delta f\,\tau_{\textrm{tar}}
+
t_m\nu_{\textrm{tar}}
\big).
\label{eq:retained_phase}
\end{equation}
Since $\alpha_R$ is held fixed under Assumption~2, the retained delay
sensitivity is determined by the baseband subcarrier offset
$\kappa_n\Delta f$, rather than by the carrier frequency. Weighting
the corresponding delay--Doppler phase sensitivities by the
sensing-design coefficient $\gamma_{r,n}$ in
\eqref{eq:gamma_r_n}, and summing over the \gls{ofdm} grid, gives
\begin{equation}
\widetilde{\mathbf{J}}_o(\mathbf{p})
\triangleq
\begin{bmatrix}
K_{\tau\tau}\displaystyle\sum_{n\in\mathcal{N}}
\kappa_n^2\gamma_{r,n}
&
-K_{\tau\nu}\displaystyle\sum_{n\in\mathcal{N}}
\kappa_n\gamma_{r,n}
\\[2pt]
-K_{\tau\nu}\displaystyle\sum_{n\in\mathcal{N}}
\kappa_n\gamma_{r,n}
&
K_{\nu\nu}\displaystyle\sum_{n\in\mathcal{N}}
\gamma_{r,n}
\end{bmatrix},
\label{eq:FIM_matrix_form}
\end{equation}
where the waveform-dependent constants, obtained from the corresponding
phase-gradient products analogously to \cite{Liu2017}, are
\begin{equation}
\big(
K_{\tau\tau},
K_{\nu\nu},
K_{\tau\nu}
\big)
=
8\pi^2
\left(
(\Delta f)^2M,\,
\sum_{m=0}^{M-1}t_m^2,\,
\Delta f\sum_{m=0}^{M-1}t_m
\right)\!\!.
\label{eq:K_kernel_def}
\end{equation}
The surrogate in \eqref{eq:FIM_matrix_form} retains the derivatives of
the explicit diagonal delay--Doppler phase, while the off-diagonal echo
components affect sensing through the allocation-dependent disturbance
term in \eqref{eq:Irn_diagonal}. Derivatives of the Doppler coupling
coefficients $c_{n,k}$ and of the disturbance statistics with respect
to the target state are not included. Accordingly,
$\widetilde{\mathbf{J}}_o(\mathbf{p})$ is used as a tractable
observation-information surrogate and is not identified with the exact
\gls{fim}. Although the centered subcarrier indices satisfy
$\sum_n\kappa_n=0$, the weighted quantity
$\sum_n\kappa_n\gamma_{r,n}$ need not vanish; hence, the
delay--Doppler cross term is retained.

The prior information in Section~\ref{subsec:prior_information} is
incorporated through the additive structure commonly used in Bayesian
information formulations \cite{XuZhang2024PriorISAC}.
Specifically,
\begin{equation}
\widetilde{\mathbf{J}}(\mathbf{p})
\triangleq
\widetilde{\mathbf{J}}_o(\mathbf{p})
+
\mathbf{J}_p,
\qquad
\widetilde{\mathbf{C}}(\mathbf{p})
\triangleq
\widetilde{\mathbf{J}}^{-1}(\mathbf{p}).
\label{eq:regularized_surrogate}
\end{equation}
Here, $\widetilde{\mathbf{C}}(\mathbf{p})$ is a
prior-regularized sensing-error surrogate used both in the sensing
constraint and in the echo-reconstruction model below; it is not
interpreted as the covariance of a particular estimator or as an exact
estimation-error bound.

Since $\widetilde{\mathbf{J}}_o(\mathbf{p})$ is a nonnegative weighted
sum of phase-gradient outer products, it is positive semidefinite.
Therefore,
\begin{equation}
\widetilde{\mathbf{J}}(\mathbf{p})
\succeq
\mathbf{J}_p
\succ
\mathbf{0},
\;
\mathbf{0}
\prec
\widetilde{\mathbf{C}}(\mathbf{p})
\preceq
\mathbf{J}_p^{-1},
\;
\operatorname{Tr}
\big(
\mathbf{W}\widetilde{\mathbf{C}}(\mathbf{p})
\big)
\leq
\Gamma_{\textrm{prior}}.
\label{eq:surrogate_bounds}
\end{equation}
Thus, the inverse in \eqref{eq:regularized_surrogate} exists for every
nonnegative power allocation without introducing an additional
regularization parameter. Under the dedicated-reference model of
Assumption~4, $\mathbf{p}_r=\mathbf{0}$ yields
$\widetilde{\mathbf{J}}_o(\mathbf{p})=\mathbf{0}$ and hence equality
in \eqref{eq:surrogate_bounds}. Consequently,
$\Gamma_{\textrm{sens}}<\Gamma_{\textrm{prior}}$ requires a nonzero
sensing contribution from the current block, whereas
$\Gamma_{\textrm{sens}}\geq\Gamma_{\textrm{prior}}$ introduces no
additional sensing restriction. This condition does not require
positive sensing power on every subcarrier and does not by itself
guarantee feasibility under the available transmit-power budget.
\vspace{-0.3cm}
% ==========================================
\subsection{Echo Reconstruction Mismatch Model}
\label{subsec:channel_error_analysis}
% ==========================================

We next relate the sensing-error surrogate
$\widetilde{\mathbf{C}}(\mathbf{p})$ to the residual echo power that
appears in \eqref{eq:mismatch_power}. Let
$\boldsymbol{\epsilon}_{\theta}
\triangleq
\boldsymbol{\theta}-\widehat{\boldsymbol{\theta}}
=
[\epsilon_\tau,\epsilon_\nu]^T$
denote the current-block target-state estimation error, which is
distinct from the pre-transmission uncertainty described by the prior
in \eqref{eq:prior_model}. With $\alpha_R$ fixed under Assumption~2,
a first-order expansion of the explicit delay--Doppler phase, while
holding the Doppler coupling coefficient $c_{n,k}$ fixed, gives

\begin{equation}
\begin{aligned}
\big[
\mathbf{H}_{\textrm{FD}}^{(\mathrm{EP})}[m]
-
\widehat{\mathbf{H}}_{\textrm{FD}}^{(\mathrm{EP})}[m]
\big]_{n,k}
&\approx
j\,
\mathbf{g}_{k,m}^{T}
\boldsymbol{\epsilon}_{\theta}
H_{\textrm{FD},n,k}^{(\mathrm{EP})}[m],
\\
\mathbf{g}_{k,m}
&\triangleq
2\pi
[-\kappa_k\Delta f,\ t_m]^T .
\end{aligned}
\label{eq:entrywise_reconstruction_error}
\end{equation}
The delay component of $\mathbf{g}_{k,m}$ is indexed by the source
subcarrier $k$, consistently with the delay phase in
\eqref{eq:h_tar_eff_definition}.

Using $\widetilde{\mathbf{C}}(\mathbf{p})$ as a proxy for the
second moment of $\boldsymbol{\epsilon}_{\theta}$, the corresponding
local reconstruction sensitivity is
\begin{equation}
\begin{aligned}
\widetilde{\Omega}[k,m]
&\triangleq
\mathbf{g}_{k,m}^{T}
\widetilde{\mathbf{C}}(\mathbf{p})
\mathbf{g}_{k,m}
\\
&=
4\pi^2
\Big[
\widetilde{C}_{\tau\tau}
(\kappa_k\Delta f)^2
+
\widetilde{C}_{\nu\nu}t_m^2
-
2\widetilde{C}_{\tau\nu}
\kappa_k\Delta f\,t_m
\Big],
\end{aligned}
\label{eq:omega_definition}
\end{equation}
where
$\widetilde{C}_{xy}
\triangleq
[\widetilde{\mathbf{C}}(\mathbf{p})]_{xy}$.
The quantity $\widetilde{\Omega}[k,m]\geq0$ therefore measures the
modeled sensitivity of the reconstructed phase on subcarrier
$k$ and symbol $m$ to target-state estimation error.

Combining this local sensitivity with the prior-average echo power,
we model the residual power of channel entry $(n,k)$ as
$\widetilde{\sigma}_{e,n,k}^{2}[m]
\triangleq
L_{n,k}\widetilde{\Omega}[k,m]$.
Since $L_{n,k}$ is independent of $m$, the blockwise mismatch model in
\eqref{eq:mismatch_power} uses

\begin{equation}
\widetilde{\Omega}_k(\mathbf{p})
\triangleq
\frac{1}{M}
\sum_{m=0}^{M-1}
\widetilde{\Omega}[k,m].
\label{eq:omega_block_average}
\end{equation}
The resulting sensitivity varies across the \gls{ofdm} grid:
subcarriers with larger $|\kappa_k|$ are more sensitive to delay error,
whereas later symbols are more sensitive to Doppler error. Moreover,
\eqref{eq:surrogate_bounds} implies
\begin{equation}
0
\leq
\widetilde{\Omega}[k,m]
\leq
4\pi^2
\left[
\sigma_\tau^2(\kappa_k\Delta f)^2
+
\sigma_\nu^2t_m^2
\right],
\label{eq:omega_bound}
\end{equation}
so the modeled sensitivity remains finite even when the current
observation contributes no sensing information.

To identify precisely which terms are affected by reconstruction
error, define
$\mathbf{u}[m]\triangleq
\mathbf{x}_r[m]+\mathbf{x}_1[m]$
as the known components whose echoes are cancelled after sensing.
Assuming correct robust-stream decoding and exact removal of the known
\gls{dp} components, subtraction of their reconstructed echoes gives
the exact post-cancellation observation
\begin{equation}
\begin{aligned}
\mathbf{y}_2[m]
&=
\big(
\mathbf{H}_{\textrm{FD}}^{(\mathrm{DP})}[m]
+
\mathbf{H}_{\textrm{FD}}^{(\mathrm{EP})}[m]
\big)
\mathbf{x}_2[m]
+
\mathbf{e}[m]
+
\mathbf{z}[m],
\\
\mathbf{e}[m]
&\triangleq
\big(
\mathbf{H}_{\textrm{FD}}^{(\mathrm{EP})}[m]
-
\widehat{\mathbf{H}}_{\textrm{FD}}^{(\mathrm{EP})}[m]
\big)
\mathbf{u}[m].
\end{aligned}
\label{eq:post_cancellation_residual}
\end{equation}
Thus, reconstruction mismatch arises only from the cancelled known components; the supplementary-stream echo remains in the received signal. Since $\widehat{\boldsymbol{\theta}}$ is estimated in the presence of $\mathbf{x}_2[m]$ and $\mathbf{z}[m]$, $\mathbf{e}[m]$ need not be independent of either.

\textit{Remark~3 (Scope of the reconstruction model):}
The mismatch power in \eqref{eq:mismatch_power} is a local design
model obtained by combining the first-order phase approximation in
\eqref{eq:entrywise_reconstruction_error}, the sensing-error surrogate
$\widetilde{\mathbf{C}}(\mathbf{p})$, and the prior-average echo power
$L_{n,k}$. It does not constitute the exact covariance of the
post-cancellation residual. In particular, perturbations of the
Doppler coupling coefficients, higher-order phase terms, residual bias,
and cross-correlations across subcarriers and with the remaining signal
and noise are not modeled. Uncertainty in $\alpha_R$ is likewise
excluded under Assumption~2.

% ==========================================
\section{Power Allocation and Proposed Solution}
\label{sec:optimisation}
% ==========================================

We optimize the per-subcarrier power allocation using the effective
sum-\gls{se} utility
\begin{equation}
\begin{aligned}
R_{\textrm{sum}}^{\textrm{eff}}(\mathbf{p})
\triangleq
\sum_{n\in\mathcal{N}}
\Big[
&\log_2\!\big(1+\gamma_{c,1,n}(\mathbf{p})\big)
\\[-1pt]
&+
\log_2\!\big(1+\gamma_{c,2,n}(\mathbf{p})\big)
\Big].
\end{aligned}
\label{eq:sum_rate_block}
\end{equation}
The \gls{sinr} terms in \eqref{eq:sum_rate_block} are constructed from
the block-averaged statistical coefficients in
\eqref{eq:block_statistical_coefficients} and the block-averaged
reconstruction sensitivities. Consequently,
\eqref{eq:sum_rate_block} is an allocation metric based on effective
blockwise \glspl{sinr}, rather than an average of symbolwise achievable
rates. The corresponding average per-subcarrier metric is
$R_{\textrm{sum}}^{\textrm{eff}}/N_{\textrm{sc}}$. 

Sensing performance is constrained through the weighted A-optimal
surrogate score
\begin{equation}
\operatorname{Tr}
\big(
\mathbf{W}\widetilde{\mathbf{C}}(\mathbf{p})
\big),
\label{eq:a_optimal_criterion}
\end{equation}
where the fixed weighting matrix $\mathbf{W}$, defined in
Section~\ref{subsec:prior_information}, places the delay and Doppler
components on a common dimensionless scale. Since
$\widetilde{\mathbf{C}}(\mathbf{p})$ is a sensing-error surrogate,
\eqref{eq:a_optimal_criterion} is likewise used as a design criterion
rather than as a physical range- or velocity-error measure.

\textit{Remark~4 (Scope of the sensing criterion):}
Frequency-dependent sensing power allocation modifies the effective
sensing spectrum and can therefore alter the delay response
\cite{Sturm2011}. The criterion in
\eqref{eq:a_optimal_criterion} captures local delay--Doppler
sensitivity, but does not characterize ambiguity sidelobes, detection
performance, or multi-target resolution.

The communication and sensing metrics depend jointly on the power
allocation, leading to
\begin{subequations}
\label{eq:raw_opt}
\begin{align}
\max_{\mathbf{p}}\quad
&
R_{\textrm{sum}}^{\textrm{eff}}(\mathbf{p})
\label{eq:raw_obj}
\\
\text{s.t.}\quad
&
\operatorname{Tr}
\big(
\mathbf{W}\widetilde{\mathbf{C}}(\mathbf{p})
\big)
\leq
\Gamma_{\textrm{sens}},
\label{eq:raw_sens_target}
\\
&
\sum_{n\in\mathcal{N}}
\big(
p_{c,1}[n]
+
p_{c,2}[n]
+
p_r[n]
\big)
\leq
P_{\textrm{tx}},
\label{eq:raw_power_tot}
\\
&
P_{\textrm{min}}
\leq
p_{c,1}[n]+p_{c,2}[n]+p_r[n]
\leq
P_{\textrm{max}},
\,
\forall n\in\mathcal{N},
\label{eq:raw_power_tube}
\\
&
p_{c,1}[n]\geq0,\,
p_{c,2}[n]\geq0,\,
p_r[n]\geq0,
\,
\forall n\in\mathcal{N}.
\label{eq:raw_nonneg}
\end{align}
\end{subequations}
Here, $\Gamma_{\textrm{sens}}$ is the maximum admissible weighted
sensing score and $P_{\textrm{tx}}$ is the transmit-power budget per
\gls{ofdm} symbol. We consider
$\Gamma_{\textrm{sens}}<\Gamma_{\textrm{prior}}$, such that prior
information alone does not satisfy the sensing requirement. The
optional bounds in \eqref{eq:raw_power_tube} limit excessive
concentration of power on individual subcarriers; they can be removed,
or equivalently relaxed by setting
$P_{\textrm{min}}=0$ and
$P_{\textrm{max}}=P_{\textrm{tx}}$.

The \gls{rs}-inspired feasible set contains the two fixed-order
architectures as special cases:
$\mathbf{p}_{c,2}=\mathbf{0}$ recovers \gls{noma}-CF, whereas
$\mathbf{p}_{c,1}=\mathbf{0}$ recovers \gls{noma}-SF. Hence, the
proposed formulation permits an interior stream split but does not
require one. More general subcarrier-dependent decoding orders would
introduce discrete order-selection variables coupled through the common
sensing and reconstruction model and are outside the present scope of this work.

Problem~\eqref{eq:raw_opt} is generally nonconvex. The communication
objective contains coupled fractional \gls{sinr} expressions, while
$\widetilde{\mathbf{C}}(\mathbf{p})$ depends nonlinearly on the
allocation and also enters the supplementary-stream mismatch through
$\widetilde{\Omega}_k(\mathbf{p})$. We address these coupled
dependencies through a sensing epigraph and successive convex
power-allocation subproblems.
\vspace{-0.3cm}
%==========================================
\subsection{A-Optimal Design and Surrogate-Based Reformulation}
%==========================================
We first express the weighted sensing constraint through an \gls{lmi}
epigraph and then isolate the allocation dependence of the
echo-reconstruction mismatch by fixing its sensitivities within each
power-update step.
\paragraph{LMI Representation of the Sensing Criterion}
Following the standard epigraph construction for A-optimal design
\cite{Joshi2009}, introduce
$\mathbf{X}\in\mathbb{S}^{2}$ and auxiliary variables
$z_n$ satisfying
$0\leq z_n\leq\gamma_{r,n}(\mathbf{p})$.
Define
\begin{equation}
\mathbf{K}_n
\triangleq
\begin{bmatrix}
K_{\tau\tau}\kappa_n^2 & -K_{\tau\nu}\kappa_n\\
-K_{\tau\nu}\kappa_n & K_{\nu\nu}
\end{bmatrix}
\succeq\mathbf{0},
\label{eq:Kn_definition}
\end{equation}
so that the observation-information surrogate becomes affine in
$\mathbf{z}$,
\begin{equation}
\widetilde{\mathbf{J}}_o(\mathbf{z})
\triangleq
\sum_{n\in\mathcal{N}}z_n\mathbf{K}_n.
\label{eq:J_affine_z}
\end{equation}
The sensing constraint can then be represented by
\begin{equation}
\begin{bmatrix}
\mathbf{X} & \mathbf{I}_2\\
\mathbf{I}_2 &
\widetilde{\mathbf{J}}_o(\mathbf{z})+\mathbf{J}_p
\end{bmatrix}
\succeq\mathbf{0},
\qquad
\operatorname{Tr}(\mathbf{W}\mathbf{X})
\leq
\Gamma_{\textrm{sens}}.
\label{eq:lmi_X_Jz}
\end{equation}
By the Schur complement,
\eqref{eq:lmi_X_Jz} implies
\[
\mathbf{X}
\succeq
\big(
\widetilde{\mathbf{J}}_o(\mathbf{z})+\mathbf{J}_p
\big)^{-1}.
\]
Moreover, since $\mathbf{K}_n\succeq\mathbf{0}$ and
$z_n\leq\gamma_{r,n}(\mathbf{p})$,
\[
\widetilde{\mathbf{J}}_o(\mathbf{z})
\preceq
\widetilde{\mathbf{J}}_o(\mathbf{p}),
\]
and hence the inverse ordering ensures that
\eqref{eq:lmi_X_Jz} implies the original sensing constraint
\eqref{eq:raw_sens_target}. Conversely, any allocation satisfying
\eqref{eq:raw_sens_target} admits
$z_n=\gamma_{r,n}(\mathbf{p})$ and
$\mathbf{X}=\widetilde{\mathbf{C}}(\mathbf{p})$.
Thus, the epigraph representation is exact when the original
fractional relations
$z_n\leq\gamma_{r,n}(\mathbf{p})$ are retained. Their tractable
inner approximation is introduced in the subsequent
subproblem.

\paragraph{Fixed-Sensitivity Approximation of the Mismatch}

The mismatch power in \eqref{eq:mismatch_power} depends on the
allocation through both the cancelled-signal powers and the
reconstruction sensitivities
$\widetilde{\Omega}_k(\mathbf{p})$.
At outer iteration $\ell$, we evaluate these sensitivities at the
preceding allocation and hold them fixed during the subsequent power
update,
\begin{equation}
\begin{aligned}
\widetilde{\Omega}_k^{(\ell)}
&\triangleq
\widetilde{\Omega}_k
\big(\mathbf{p}^{(\ell-1)}\big),
\\
\widetilde{P}_{\textrm{mismatch},n}^{(\ell)}(\mathbf{p})
&\triangleq
\sum_{k\in\mathcal{N}}
\widetilde{\Omega}_k^{(\ell)}
L_{n,k}
\big(
p_{c,1}[k]+p_r[k]
\big).
\end{aligned}
\label{eq:frozen_sensitivity_update}
\end{equation}
The full matrix
$\widetilde{\mathbf{C}}
(\mathbf{p}^{(\ell-1)})$,
including its delay--Doppler cross term, is retained when computing
$\widetilde{\Omega}_k^{(\ell)}$.
For fixed sensitivities,
$\widetilde{P}_{\textrm{mismatch},n}^{(\ell)}(\mathbf{p})$
is affine and nonnegative in the power variables, which enables the
convex power-update construction below.
\par Freezing the sensitivities does not, however, provide a global upper
or lower bound on the original objective. Accordingly, after each
candidate update, the sensing surrogate, reconstruction sensitivities,
and effective sum-\gls{se} are recomputed using the candidate
allocation before it is accepted.
%==========================================
\subsection{Fractional Programming and Dual Transformations}
\label{subsec:fractional_programming}
%==========================================

With the reconstruction sensitivities fixed as in
\eqref{eq:frozen_sensitivity_update}, the communication \gls{sinr}
numerators and denominators are affine functions of the power
allocation. We therefore apply the fractional-programming transforms
of \cite{Shen2018} to obtain concave communication surrogates and a
convex inner approximation of the sensing links within each power
update.

%------------------------------------------
\paragraph{Lagrangian Dual Transform for the Communication Utility}
%------------------------------------------

For each communication stream $i\in\{1,2\}$ and subcarrier $n$, the
Lagrangian dual transform introduces an auxiliary variable
$\alpha_{c,i,n}\geq0$ such that \cite{Shen2018}
\begin{equation}
\begin{aligned}
\log_2\!\big(1+\gamma_{c,i,n}\big)
=\!\!\!
\max_{\alpha_{c,i,n}\geq0}
\Bigg\{&\!
\log_2\!\big(1+\alpha_{c,i,n}\big)
-\alpha_{c,i,n}\log_2 e
\\
&+
\frac{(1+\alpha_{c,i,n})\gamma_{c,i,n}}
{1+\gamma_{c,i,n}}
\log_2 e
\Bigg\},
\end{aligned}
\label{eq:dual_transform_time}
\end{equation}
with maximizer
$\alpha_{c,i,n}^{\star}=\gamma_{c,i,n}$.
For the supplementary stream, the denominator is evaluated using the
fixed reconstruction sensitivities of
\eqref{eq:frozen_sensitivity_update}.

%------------------------------------------
\paragraph{Quadratic Transform for the Communication Ratios}
%------------------------------------------

Under the fixed-sensitivity approximation, define the supplementary
stream disturbance as
\begin{equation}
\begin{aligned}
I_{c,2,n}^{(\ell)}(\mathbf{p})
\triangleq\,
&\mathcal{V}_np_{c,2}[n]
+
\!\!\!\!\!\!\\
&
\sum_{k\in\mathcal{N}\setminus\{n\}}
\!\!\!\!\!\!
L_{n,k}p_{c,2}[k]
+
\widetilde{P}_{\textrm{mismatch},n}^{(\ell)}(\mathbf{p}),
\end{aligned}
\label{eq:Ic2_def_fp}
\end{equation}
which is affine in $\mathbf{p}$. For notational consistency, let
$I_{c,1,n}^{(\ell)}(\mathbf{p})\equiv I_{c,1,n}(\mathbf{p})$.
For $i\in\{1,2\}$, define
\begin{equation}
\begin{aligned}
A_{i,n}
&\triangleq
\log_2(1+\alpha_{c,i,n})
-\alpha_{c,i,n}\log_2 e,\\
B_{i,n}
&\triangleq
\sqrt{(1+\alpha_{c,i,n})S_{c,i,n}},\\
C_{i,n}
&\triangleq
S_{c,i,n}
+
I_{c,i,n}^{(\ell)}(\mathbf{p})
+
\sigma_{\textrm{BS}}^2.
\end{aligned}
\label{eq:ABC_defs}
\end{equation}
Introducing $\rho_{c,i,n}\geq0$, the quadratic transform gives
\begin{equation}
\widetilde{R}_{c,i,n}
\triangleq
A_{i,n}
+
\Big(
2\rho_{c,i,n}B_{i,n}
-
\rho_{c,i,n}^{2}C_{i,n}
\Big)\log_2 e .
\label{eq:R_ci_surrogate}
\end{equation}
For fixed auxiliaries and reconstruction sensitivities,
$\widetilde{R}_{c,i,n}$ is concave in the power variables. For fixed
powers and $\alpha_{c,i,n}$, its maximizing quadratic-transform
auxiliary is \cite{Shen2018}
\begin{equation}
\rho_{c,i,n}^{\star}
=
\frac{B_{i,n}}{C_{i,n}},
\qquad
i\in\{1,2\}.
\label{eq:rho_update}
\end{equation}
Maximization over both auxiliary variables recovers the corresponding
communication utility for fixed reconstruction sensitivities. When
the auxiliaries are instead fixed at their maximizing values for the
preceding allocation, the transformed expression is a concave lower
bound that is tight at that allocation. This lower-bound property
applies to the fixed-sensitivity utility and not, in general, to the
full objective with allocation-dependent reconstruction sensitivities.

%------------------------------------------
\paragraph{Quadratic Transform for the Sensing Links}
%------------------------------------------

The fractional sensing links
$z_n\leq\gamma_{r,n}(\mathbf{p})$ are handled similarly. At iteration
$\ell$, fix $y_n^{(\ell)}\geq0$ and impose
\begin{equation}
\begin{aligned}
z_n
\leq\,
&2y_n^{(\ell)}
\sqrt{S_{r,n}(\mathbf{p})}
-
\big(y_n^{(\ell)}\big)^2
\big(
I_{r,n}(\mathbf{p})
+
\sigma_{\textrm{BS}}^2
\big),
\,
\end{aligned}
\label{eq:fp_quadratic_transform_sensing_revised}
\end{equation}
$\forall n\in\mathcal{N}$. For fixed $y_n^{(\ell)}$, the right-hand side is concave in
$\mathbf{p}$ and lower-bounds
$\gamma_{r,n}(\mathbf{p})$. Equality holds at the preceding
allocation when \cite{Shen2018}
\begin{equation}
y_n^{(\ell)}
=
\frac{
\sqrt{S_{r,n}(\mathbf{p}^{(\ell-1)})}
}{
I_{r,n}(\mathbf{p}^{(\ell-1)})
+
\sigma_{\textrm{BS}}^2
}.
\label{eq:sensing_auxiliary_update}
\end{equation}
Together with $z_n\geq0$ and the \gls{lmi} in
\eqref{eq:lmi_X_Jz}, this provides a convex inner approximation of the
sensing constraint that is tight at the preceding allocation.

%------------------------------------------
\paragraph{Convex Power-Allocation Update}
%------------------------------------------

For fixed communication auxiliaries
$\{\boldsymbol{\alpha}^{(\ell)},\boldsymbol{\rho}^{(\ell)}\}$,
sensing auxiliaries $\mathbf{y}^{(\ell)}$, and reconstruction
sensitivities
$\{\widetilde{\Omega}_k^{(\ell)}\}_{k\in\mathcal{N}}$,
the next power-allocation candidate is obtained by solving
\begin{subequations}
\label{eq:maxSE_hard}
\begin{align}
\max_{\mathbf{p},\,\mathbf{X},\,\mathbf{z}}
\quad &
\sum_{n\in\mathcal{N}}
\Big(
\widetilde{R}_{c,1,n}^{(\ell)}(\mathbf{p})
+
\widetilde{R}_{c,2,n}^{(\ell)}(\mathbf{p})
\Big)
\\
\text{s.t.}\quad
&
\eqref{eq:lmi_X_Jz},
\qquad
\mathbf{z}\succeq\mathbf{0},
\\
&
\eqref{eq:fp_quadratic_transform_sensing_revised}
\ \text{with}\ 
y_n=y_n^{(\ell)},
\\
&
\eqref{eq:raw_power_tot},\
\eqref{eq:raw_power_tube},\
\eqref{eq:raw_nonneg}.
\end{align}
\end{subequations}
Problem~\eqref{eq:maxSE_hard} is a convex conic program containing a
single semidefinite constraint. The resulting candidate is
subsequently evaluated using the allocation-dependent reconstruction
sensitivities and the original objective, as described in
Section~\ref{subsec:algorithm}.
\paragraph{Numerical Scaling}
For numerical conditioning, powers are normalized by the average
available subcarrier power $P_{\textrm{avg}}$, and the effective
channel coefficients are scaled by
$\sqrt{P_{\textrm{avg}}/\sigma_{\textrm{BS}}^2}$, yielding unit
receiver-noise variance. For the sensing \gls{lmi}, define
$\mathbf{S}_p\triangleq
\operatorname{diag}(\sigma_\tau,\sigma_\nu)$ and apply the congruence
scaling
\begin{equation}
\mathbf{X}_{s}
=
\mathbf{S}_p^{-1}\mathbf{X}\mathbf{S}_p^{-1},
\quad
\widetilde{\mathbf{J}}_{s}(\mathbf{z})
=
\mathbf{I}_2
+
\sum_{n\in\mathcal{N}}
z_n\mathbf{S}_p\mathbf{K}_n\mathbf{S}_p .
\label{eq:normalized_epigraph_variables}
\end{equation}
This maps the prior information matrix to the identity while
preserving the \gls{lmi} structure and the feasible power allocations.
A further fixed diagonal congruence is applied within each iteration
to balance the numerical scale of the two sensing dimensions; this
scaling affects conditioning only and does not alter the feasible set.
\vspace{-0.5cm}
%==========================================
\subsection{Implementation, Initialization, and Computational Cost}
\label{subsec:algorithm}
%==========================================
Algorithm~\ref{alg:bcd} summarizes the proposed power-allocation
procedure. Each start alternates closed-form updates of the
fractional-programming auxiliaries and reconstruction sensitivities
with the convex power update in \eqref{eq:maxSE_hard}. Since the
fixed-sensitivity approximation does not globally bound the original
objective, every candidate update is subsequently evaluated using the
full allocation-dependent sensing and reconstruction model and is
accepted only if it remains feasible and improves the effective
sum-\gls{se}.
\begin{algorithm}[!htb]
\caption{\bc{Sensing-Constrained Power Allocation}}
\color{black}
\label{alg:bcd}
\begin{algorithmic}[1]

\State \textbf{Input:}
$\mathcal{I}^{-}$, $\Gamma_{\textrm{sens}}$, power constraints,
architecture, candidate set $\mathcal{S}$, $N_{\textrm{st}}$,
$I_{\max}$, tolerances, and
$\mathcal{B}=\{2^{-j}\}_{j=0}^{7}$

\State \textbf{Output:}
feasible allocation $\mathbf{p}^{\star}$ or initialization-failure status

\State Compute $\mathbf{L}$ and
$\{\mathcal{G}_n,\mathcal{V}_n\}_{n\in\mathcal{N}}$
from $\mathcal{I}^{-}$

\State Audit $\mathcal{S}$ and discard infeasible candidates;
\textbf{terminate} if none remain. Otherwise, sort by decreasing
$R_{\textrm{sum}}^{\textrm{eff}}$, set
$\mathbf{p}^{\star}=\mathbf{p}_1$, and
$J=\min\{N_{\textrm{st}},|\mathcal{S}|\}$

\State For $j=2,\ldots,J$, replace
$\mathbf{p}_j$ by
$0.98\,\mathbf{p}_j+0.02\,\mathbf{p}^{\star}$
if the resulting point is feasible
\label{alg:line_mix}

\For{$j=1,\ldots,J$, with $\mathbf{p}^{(0)}=\mathbf{p}_j$}

    \For{$\ell=1,\ldots,I_{\max}$}

        \State At $\mathbf{p}^{(\ell-1)}$, compute
        $\widetilde{\mathbf{C}}$,
        $\{\widetilde{\Omega}_k^{(\ell)}\}_{k\in\mathcal{N}}$,
        and $\mathbf{g}^{(\ell)}$; update
        $\boldsymbol{\alpha}$, $\boldsymbol{\rho}$, and $\mathbf{y}$
        in closed form

        \State Solve \eqref{eq:maxSE_hard} with the objective augmented
        by \eqref{eq:chain_correction}, yielding
        $\widehat{\mathbf{p}}^{(\ell)}$;
        \textbf{break} if no acceptable solver solution is obtained
        \label{alg:line_solve}

        \State Project numerical negative entries of
        $\widehat{\mathbf{p}}^{(\ell)}$ to zero and enforce the
        architecture-specific inactive stream;
        \textbf{break} if the required repair exceeds
        $\epsilon_{\textrm{rep}}$

        \State Audit
        $\mathbf{p}_{\mu}
        =
        \mathbf{p}^{(\ell-1)}
        +
        \mu\big(
        \widehat{\mathbf{p}}^{(\ell)}
        -
        \mathbf{p}^{(\ell-1)}
        \big)$
        for $\mu\in\mathcal{B}$; select the feasible trial maximizing
        $R_{\textrm{sum}}^{\textrm{eff}}$;
        \textbf{break} if no trial improves the current objective
        \label{alg:line_audit}

        \State Set $\mathbf{p}^{(\ell)}$ to the selected trial and
        update $\mathbf{p}^{\star}$ if the incumbent objective improves

        \State \textbf{break} if $\ell\geq3$ and both stopping
        criteria are satisfied

    \EndFor
\EndFor
\State \textbf{return} $\mathbf{p}^{\star}$
\end{algorithmic}
\end{algorithm}
\paragraph{Initialization and Endpoint Preservation}
A deterministic feasibility-screened candidate set $\mathcal{S}$ is
constructed from a collection of sensing-power profiles and
communication-stream splits, without invoking the convex solver.
Feasible candidates are ranked according to
$R_{\textrm{sum}}^{\textrm{eff}}$, and up to
$N_{\textrm{st}}$ candidates with the largest objective values are used
to initialize the multi-start procedure. The specific candidate
profiles and numerical search parameters are given in
Section~\ref{sec:numerical_results}.

For the \gls{rs}-inspired design, $\mathcal{S}$ additionally includes
the feasible solutions returned by the \gls{noma}-CF and
\gls{noma}-SF procedures. Since the incumbent
$\mathbf{p}^{\star}$ is initialized with the best feasible candidate
and subsequently replaced only by feasible objective-improving
updates,
\begin{equation}
R_{\textrm{RS}}
\geq
\max\{R_{\textrm{CF}},R_{\textrm{SF}}\}
\label{eq:endpoint_preservation}
\end{equation}
holds relative to the returned fixed-order solutions by construction.
This endpoint-preservation property does not imply global optimality.
The small feasible mixture in line~\ref{alg:line_mix} is used only to
perturb secondary initializations away from degenerate zero-power
components.

\paragraph{First-Order Sensitivity Correction}
Freezing $\widetilde{\Omega}_k$ omits the dependence of the objective
on $\mathbf{p}$ through the reconstruction sensitivities. We account
for this dependence locally by augmenting the power-update objective
with the first-order term
$(\mathbf{g}^{(\ell)})^T
(\mathbf{p}-\mathbf{p}^{(\ell-1)})$, where
\begin{equation}
\mathbf{g}^{(\ell)}
\triangleq
\left.
\sum_{k\in\mathcal{N}}
\frac{\partial R_{\textrm{sum}}^{\textrm{eff}}}
{\partial\widetilde{\Omega}_k}
\nabla_{\mathbf{p}}\widetilde{\Omega}_k(\mathbf{p})
\right|_{\mathbf{p}=\mathbf{p}^{(\ell-1)}} .
\label{eq:chain_correction}
\end{equation}
The required gradient follows from
$\mathrm{d}\widetilde{\mathbf{C}}
=-\widetilde{\mathbf{C}}\,
\mathrm{d}\widetilde{\mathbf{J}}\,
\widetilde{\mathbf{C}}$
and \eqref{eq:omega_definition}. The correction restores the omitted
chain-rule contribution at the current iterate but does not constitute
a global lower bound; candidate improvement is therefore assessed
using the recomputed objective.

\paragraph{Numerical Solution, Acceptance, and Stopping}
The convex subproblem is solved using CVX with the numerical scaling
described in Section~\ref{subsec:fractional_programming}. For the two
fixed-order baselines, the inactive communication stream and its
identically zero fractional-programming terms are removed from the
subproblem. Each candidate generated in line~\ref{alg:line_audit} is evaluated
using the original allocation-dependent model: the sensing weights,
$\widetilde{\mathbf{C}}(\mathbf{p})$, reconstruction sensitivities,
and $R_{\textrm{sum}}^{\textrm{eff}}(\mathbf{p})$ are recomputed from
the candidate powers, and the original design constraints are checked.
A candidate is accepted only if it is feasible and improves the
recomputed effective sum-\gls{se}. Consequently, the accepted objective
values are nondecreasing within each start. A start terminates upon solver failure, excessive numerical repair,
absence of an improving feasible trial, attainment of $I_{\max}$, or
simultaneous satisfaction of the objective- and allocation-change
criteria
$\Delta R^{(\ell)}
\leq
\epsilon_R\max\{1,R_{\textrm{sum}}^{\textrm{eff}}
(\mathbf{p}^{(\ell)})\}$
and
$\|\mathbf{p}^{(\ell)}-\mathbf{p}^{(\ell-1)}\|_F/
\max\{1,\|\mathbf{p}^{(\ell-1)}\|_F\}
\leq\epsilon_p$.
\bc{Power-change and numerical-repair tests use powers
normalized by $P_{\textrm{avg}}$. CVX uses {high} precision; stopping and repair tolerances
are specified in Section~\ref{sec:numerical_results}.} The procedure does not imply
stationarity or global optimality for \eqref{eq:raw_opt}.

\paragraph{Computational Cost}
For the \gls{rs}-inspired design, each convex power-update problem
contains $4N_{\textrm{sc}}+3$ scalar decision variables, compared with
$3N_{\textrm{sc}}+3$ for either fixed-order baseline, together with one
$4\times4$ semidefinite constraint and $O(N_{\textrm{sc}})$ additional
convex constraints. Each outer iteration evaluates the reconstruction
sensitivities and the correction in \eqref{eq:chain_correction}, solves
the convex subproblem, and evaluates the trial allocations used in the
acceptance step. The latter requires
$O(N_{\textrm{sc}}^2)$ echo-\gls{ici} operations and
$O(MN_{\textrm{sc}})$ sensitivity operations per trial. An indicative
dense interior-point scaling therefore gives
\begin{equation}
\mathcal{C}_{\mathrm{it}}
=
O\!\left(
|\mathcal{B}|
\big(
N_{\textrm{sc}}^2+MN_{\textrm{sc}}
\big)
\right)
+
O\!\left(
N_{\mathrm{it}}^{\mathrm{ip}}N_{\textrm{sc}}^3
\right),
\label{eq:complexity_summary}
\end{equation}
where $N_{\mathrm{it}}^{\mathrm{ip}}$ denotes the number of
interior-point iterations. This expression is intended to indicate the
dominant scaling rather than a solver-specific worst-case bound.

The prior-dependent quantities are computed once per block.
Constructing the circulant matrix $\mathbf{L}$ requires
$O(N_{\textrm{sc}}^2)$ operations, while evaluating the diagonal prior
echo mean over the $M\times N_{\textrm{sc}}$ resource grid requires
$O(MN_{\textrm{sc}}^2)$. The candidate-generation stage invokes no
convex solver, and each architecture performs at most
$N_{\textrm{st}}I_{\max}$ outer iterations across all starts. The
number of optimization variables and fractional-programming
auxiliaries is independent of $M$.
\vspace{-0.2cm}

\section{Numerical Results}
\label{sec:numerical_results}
This section evaluates the proposed \gls{rs}-inspired bistatic
\gls{ofdm}-\gls{isac} design against the two fixed-order architectures
and examines its power-allocation behavior across different sensing and
echo-channel conditions.
\vspace{-0.2cm}
\subsection{Simulation Setup}
We consider blocks of $N_{\textrm{sc}}=32$ subcarriers with subcarrier
spacing $\Delta f=15$~kHz and $M=16$ \gls{ofdm} symbols. Performance is
reported in terms of the average effective \gls{se}
$R_{\textrm{sum}}^{\textrm{eff}}/N_{\textrm{sc}}$ defined in
\eqref{eq:sum_rate_block}. Consistent with the formulation, this metric
excludes \gls{cp}, acquisition, pilot, and control overhead and
characterizes the adopted statistical communication and sensing model
rather than coded throughput.

The proposed design jointly allocates power to the robust,
supplementary, and sensing components. It is compared with the two
fixed-order endpoints: \gls{noma}-CF, obtained by setting
$\mathbf{p}_{c,2}=\mathbf{0}$, and \gls{noma}-SF, by setting
$\mathbf{p}_{c,1}=\mathbf{0}$. For each channel realization, all three
architectures use the same channel, target-state prior, sensing
requirement, and power constraints. Denoting their effective sum
\glspl{se} by $R_{\textrm{CF},i}$,
$R_{\textrm{SF},i}$, and $R_{\textrm{RS},i}$, respectively, we define
the fixed-order envelope as
$R_{\textrm{env},i}\triangleq
\max\{R_{\textrm{CF},i},R_{\textrm{SF},i}\}$.
\bc{The realization-wise percentage gain is
$G_i\triangleq100(R_{\textrm{RS},i}/R_{\textrm{env},i}-1)$.}
Since the proposed algorithm retains the returned endpoint solutions
as feasible candidates, $G_i\geq0$ follows by construction. \bc{The numerical study therefore quantifies strict gains over the returned fixed-order solutions and examines the associated communication-stream power splits.}

The sensing requirement is calibrated separately for each realization
using a radar-only reference allocation $\mathbf{p}_{\textrm{ref}}$,
for which $p_r[n]=P_{\textrm{avg}}$ and
$p_{c,1}[n]=p_{c,2}[n]=0$ for all $n$, with
$P_{\textrm{avg}}\triangleq
P_{\textrm{tx}}/N_{\textrm{sc}}$. Its weighted sensing score is
\begin{align}
S_{\textrm{ref},i}
&\triangleq
\operatorname{Tr}
\left(
\mathbf{W}
\widetilde{\mathbf{C}}_i(\mathbf{p}_{\textrm{ref}})
\right),
\label{eq:numerical_sensing_reference}\\
\Gamma_{\textrm{sens},i}(\eta)
&\triangleq
S_{\textrm{ref},i}
+
\eta
\big(
\Gamma_{\textrm{prior}}
-
S_{\textrm{ref},i}
\big),
\;
0<\eta<1 .
\label{eq:numerical_sensing_calibration}
\end{align}
For $S_{\textrm{ref},i}<\Gamma_{\textrm{prior}}$, decreasing $\eta$
tightens the sensing requirement toward the radar-only reference,
whereas $\eta\rightarrow1$ approaches the prior-only score. Unless
otherwise stated, $\eta=0.10$. We use
$\sigma_\tau=10$~ns and $\sigma_\nu=100$~Hz, together with
$\tau_{\textrm{ref}}=10$~ns and
$\nu_{\textrm{ref}}=100$~Hz, yielding
$\Gamma_{\textrm{prior}}=2$. The same value of $\eta$ is applied to all
three schemes for a given realization. Since
$S_{\textrm{ref},i}$ depends on the channel and target condition, the
resulting absolute threshold may vary across realizations while
preserving the same relative sensing calibration. In particular, a
zero prior-mean Doppler retains $\sigma_\nu=100$~Hz and therefore does
not imply a diagonal echo channel.

Unless otherwise stated, statistical comparisons use the same
$200$ channel seeds for all three architectures at each operating
condition. Matching seeds are also retained across Doppler conditions,
so that the underlying direct-path realization and echo coefficient
remain unchanged. Feasibility is checked independently for every
returned allocation, and no realization is discarded on the basis of
its achieved objective value.

The direct path is generated using independent Rayleigh taps with the
normalized TDL-C power-delay profile \cite{3gpp_38901} and total average
power
$\sum_{l=1}^{L_{\textrm{p}}}\mathbb{E}[|\alpha_l|^2]
=
10^{25/10}
(\lambda/(4\pi d_{\textrm{UE-BS}}))^2$,
where the aggregate direct-path antenna gain is included in the link
budget. The echo coefficient follows the Rician model in
\eqref{eq:bistatic_reflection_coeff}. The geometry is fixed within each
simulation campaign, and the bulk direct-path Doppler is perfectly
compensated as assumed in Section~\ref{subsec:dp_signal_model}.
Table~\ref{tab:sim_parameters} summarizes the principal simulation and
optimization parameters.
\begin{table}[!htbp]
\color{black}
\centering
\caption{\bc{Simulation Parameters}}
\label{tab:sim_parameters}
\footnotesize
\setlength{\tabcolsep}{3pt}
\begin{tabular}{@{}ll@{}}
\toprule
\textbf{Parameter} & \textbf{Value} \\
\midrule
\multicolumn{2}{@{}l}{\textit{Waveform and transmission}} \\
Carrier frequency & $28$ GHz \\
Subcarrier spacing & $15$ kHz \\
Subcarriers / symbols per block & $32$ / $16$ \\
Cyclic-prefix samples / duration & $8$ / $16.67~\mu$s \\
Transmit power budget & $23$ dBm \\
Noise power per subcarrier & $-108$ dBm \\
\midrule
\multicolumn{2}{@{}l}{\textit{Geometry and channels}} \\
\gls{bs} / \gls{ue} positions (m) & $(0,0)$ / $(300,0)$ \\
Target reference position (m) & $(250,50)$ \\
Aggregate direct-path antenna gain & $25$ dB \\
Aggregate radar antenna gain & $38$ dB \\
Direct-path profile & TDL-C \\
Delay-spread scaling parameter & $1000$ ns \\
Echo Rician factor & $10$ dB \\
\midrule
\multicolumn{2}{@{}l}{\textit{Prior and sensing requirement}} \\
Prior widths, $\sigma_\tau$, $\sigma_\nu$ & $10$ ns, $100$ Hz \\
Reference scales, $\tau_{\textrm{ref}}$, $\nu_{\textrm{ref}}$
& $10$ ns, $100$ Hz \\
Prior-only score, $\Gamma_{\textrm{prior}}$ & $2$ \\
Calibration parameter, $\eta$ & $0.10$ \\
\midrule
\multicolumn{2}{@{}l}{\textit{Power allocation and optimization}} \\
Per-subcarrier power bounds
& $[0.95,1.05]P_{\textrm{avg}}$ \\
Maximum starts / iterations per start
& $4$ / $60$ \\
Stopping tolerances, $\epsilon_R$, $\epsilon_p$
& $10^{-5}$ / $10^{-4}$ \\
Line-search step set
& $\{2^{-j}\}_{j=0}^{7}$ \\
Primary solver
& CVX with SDPT3 \\
\bottomrule
\end{tabular}
\end{table}

For initialization, the deterministic candidate search described in
Section~\ref{subsec:algorithm} uses communication split parameters
$s\in\{0.2,0.5,0.8\}$ for the proposed design, together with the two
fixed-order endpoints. \bc{The first family sets
$p_r[n]=P_{\textrm{avg}}\min\{1,\zeta w_n\}$,
with $\zeta\in\{0,0.05,\ldots,1\}$ and either
$w_n=1$ or $w_n=0.5+|\kappa_n|/\max_k|\kappa_k|$.
The second assigns $P_{\textrm{avg}}$ exclusively
to sensing on progressively larger prefixes of
$(0,N_{\textrm{sc}}-1,1,N_{\textrm{sc}}-2,\ldots)$,
stopping at the first prefix admitting a feasible split.
In both families, the remaining power
$P_{\textrm{avg}}-p_r[n]$ is divided between the robust
and supplementary streams in proportions $s$ and $1-s$.} Secondary initializations use the
$0.98/0.02$ feasible mixture in Algorithm~\ref{alg:bcd}, and the
numerical repair tolerance is
$\epsilon_{\textrm{rep}}=10^{-5}$.
\subsection{Iterative Behavior}
Figure~\ref{fig:convergence_plot} shows the mean effective \gls{se} of
the accepted iterates over the same $200$ channel realizations for each
architecture, with target \gls{rcs} $10$~dBsm, mean Doppler $3$~kHz,
and $\eta=0.10$. For each realization, the trajectory corresponding to
the start attaining the largest final objective is reported. The first
$25$ iterations are shown, with the terminal value retained for starts
that terminate earlier. The shaded regions indicate pointwise $95\%$
confidence intervals for the mean.

% Requires: \usepackage{pgfplots} \usepgfplotslibrary{fillbetween}
\begin{figure}[!htbp]
\centering
\includegraphics[width=\columnwidth]{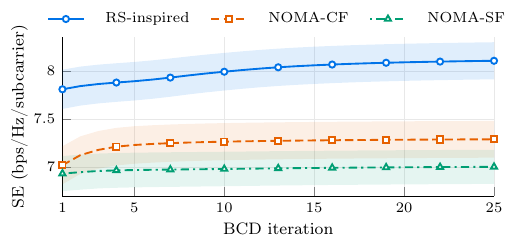}
\caption{\bc{Mean effective \gls{se} over $200$ channel seeds versus outer iteration, with pointwise $95\%$ confidence intervals.}}
\vspace{-0.3cm}
\label{fig:convergence_plot}
\end{figure}

As imposed by the acceptance step, the mean objective is nondecreasing
for all three architectures. The largest improvements occur during the
early iterations, followed by progressively smaller changes. \bc{The nearly flat \gls{noma}-SF curve is consistent with limited improvement beyond its selected initialization, whereas the \gls{rs}-inspired design additionally refines the communication-stream split during iteration.} At
iteration $25$, the mean effective \gls{se} is $8.108$, $7.291$, and
$7.004$~bps/Hz/subcarrier for the \gls{rs}-inspired, \gls{noma}-CF,
and \gls{noma}-SF designs, respectively. Between iterations $20$ and
$25$, the corresponding increases are only $0.0161$, $0.0044$, and
$0.0051$~bps/Hz/subcarrier. These results indicate diminishing
improvements with iteration, but do not imply stationarity or
convergence within the displayed horizon.

% ============================================================
% Figura 4: declarada depois, em uma coluna
% ============================================================
%\addtocounter{figure}{-2}
\begin{figure}[htbp]
    \centering
\includegraphics[width=\columnwidth]{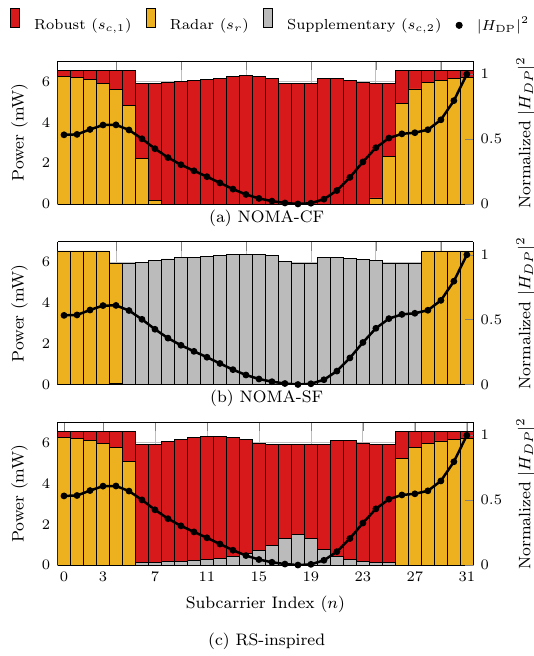}
\vspace{-0.2cm}
    \caption{Illustrative per-subcarrier power allocations for
(a) \gls{noma}-CF, (b) \gls{noma}-SF, and
(c) the \gls{rs}-inspired design. The black curve shows the normalized
direct-path power gain.}
    \vspace{-0.3cm}
\label{fig:comparison_power_allocation}
\end{figure}
%\addtocounter{figure}{1}

% Figure 5
%\addtocounter{figure}{1}
\begin{figure*}[!t]
\centering
\begin{subfigure}{0.32\textwidth}
    \centering
    \includegraphics[width=\linewidth]{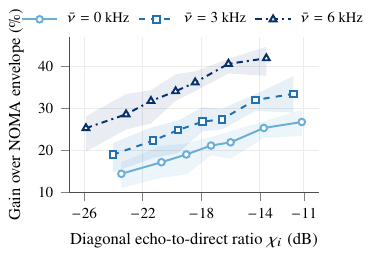}
    \caption{\bc{Envelope gain, target \gls{rcs} $15$ dBsm.}}
    \label{fig:gain_max_envelope}
\end{subfigure}
\hfill
\begin{subfigure}{0.32\textwidth}
    \centering
    \includegraphics[width=\linewidth]{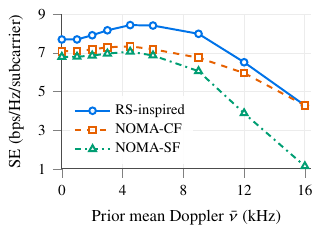}
    \caption{\bc{Effective \gls{se}, target \gls{rcs} $10$ dBsm.}}
    \label{fig:gain_vs_doppler}
\end{subfigure}
\hfill
\begin{subfigure}{0.32\textwidth}
    \centering
    \makebox[\linewidth][c]{%
    \includegraphics[width=1.2\linewidth]{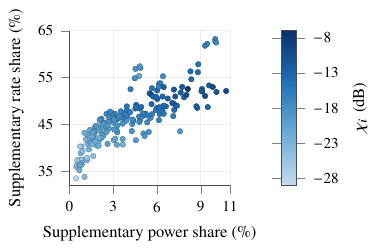}%
}
    \caption{\bc{Supplementary shares, \bc{target \gls{rcs} $15$~dBsm}.}}
    \label{fig:power_rate_shares}
\end{subfigure}
\caption{\bc{Performance over $200$ common seeds per condition at $\eta=0.10$: (a) gain over the fixed-order envelope versus echo-to-direct ratio; (b) effective \gls{se} versus Doppler; (c) supplementary-stream power and rate shares.}}
\vspace{-0.3cm}
\label{fig:channel_doppler}
\end{figure*}

\subsection{\bc{Power Allocation and Sensing--Communication Coupling}}
Figure~\ref{fig:comparison_power_allocation} illustrates the optimized
power allocations for one realization, with target
\gls{rcs} $10$~dBsm, mean Doppler $3$~kHz, and $\eta=0.05$.
The black curve shows the direct-path power gain normalized by its
maximum. This example highlights the different sensing costs associated
with the two communication streams. The robust stream is decoded before
sensing and its direct-path component is cancelled, whereas the
supplementary stream remains present during sensing. Consequently, the
two streams need not have the same preferred spectral allocation.

Across the three architectures, sensing power is concentrated toward
the band edges, consistent with the larger delay-information weights at
larger $|\kappa_n|$. For \gls{noma}-CF, robust communication and
sensing power coexist on the edge subcarriers because the robust-stream
direct-path contribution has been removed before sensing. In contrast,
\gls{noma}-SF exhibits stronger spectral separation between sensing and
supplementary communication, whose uncancelled contribution remains a
source of sensing interference.

%figure

The \gls{rs}-inspired solution combines these two allocation patterns.
The robust stream occupies much of the band and coexists with sensing
toward the edges, while the supplementary stream is concentrated in
the interior and becomes strongest around the deep direct-path fade.
This behavior is consistent with the supplementary stream being less
costly to the sensing stage on subcarriers with weaker direct-path
power. For this realization, the \gls{rs}-inspired solution allocates
$59.40\%$, $5.12\%$, and $35.47\%$ of the total power to the robust,
supplementary, and sensing components, respectively, demonstrating a
non-degenerate use of both communication processing stages under the
same sensing requirement.
\subsection{Performance across Channel and Doppler Conditions}

Figure~\ref{fig:channel_doppler} examines how the benefit of message
splitting varies with the echo-to-direct-path balance and target
Doppler, and how this benefit is reflected in the communication-stream
split. We characterize the diagonal echo-to-direct power ratio as
\begin{equation}
\chi_i \triangleq
10\log_{10}\!\left(
\frac{\sum_{n\in\mathcal{N}}L_{n,n}^{(i)}}
{\sum_{n\in\mathcal{N}}|H_{\textrm{DP},n}^{(i)}|^2}
\right).
\label{eq:echo_direct_ratio}
\end{equation}
Panel~(a) reports the median realization-wise gain over the fixed-order
envelope across seven approximately equal-count $\chi_i$ bins for mean
Dopplers of $0$, $3$, and $6$~kHz. Panel~(b) shows the median effective
\gls{se} of all three architectures over the wider Doppler range
$0$--$16$~kHz. Panel~(c) relates the supplementary stream's share of
communication power to its share of the effective communication rate
for the $200$ \gls{rs}-inspired realizations at mean Doppler $3$~kHz.
All results use $\eta=0.10$.

In Fig.~\ref{fig:channel_doppler}(a), the median gain increases across
the displayed $\chi_i$ range from $14.36\%$ to $26.75\%$ at
$0$~kHz, from $18.93\%$ to $33.46\%$ at $3$~kHz, and from
$25.27\%$ to $41.99\%$ at $6$~kHz. Thus, message splitting becomes
increasingly beneficial as the diagonal echo strengthens relative to
the direct path. The separation between the Doppler curves further
shows that $\chi_i$ alone does not determine the gain: increasing
Doppler also redistributes echo power into off-diagonal \gls{ici},
altering the interference seen across the two processing stages.

Figure~\ref{fig:channel_doppler}(b) shows the corresponding behavior
over a wider Doppler range. The \gls{rs}-inspired design achieves the
largest median effective \gls{se} among the three architectures
throughout the sampled conditions, reaching
$8.44$~bps/Hz/subcarrier at $4.5$~kHz. Its realization-wise gain over
the fixed-order envelope instead peaks at $18.46\%$ at $9$~kHz, where
its median effective \gls{se} is $7.99$~bps/Hz/subcarrier. Hence, the
Doppler condition yielding the largest relative benefit from message
splitting need not coincide with that yielding the largest absolute
communication performance.

At higher Doppler, the median effective \gls{se} decreases for all
three architectures, most sharply for \gls{noma}-SF. At $16$~kHz,
the proposed design and \gls{noma}-CF both achieve approximately
$4.28$~bps/Hz/subcarrier, compared with $1.14$ for \gls{noma}-SF.
This transition is consistent with the diagonal echo fraction
$\Lambda(0)$ decreasing from approximately $0.88$ at $3$~kHz to
$0.255$ at $9$~kHz and $4\times10^{-3}$ at $16$~kHz. As the echo
becomes increasingly off-diagonal, the contribution captured by the
diagonal sensing surrogate diminishes, and the optimized
\gls{rs}-inspired solution increasingly approaches the
communication-first endpoint.

Figure~\ref{fig:channel_doppler}(c) further shows that the resulting
communication split is strongly asymmetric. The supplementary stream
receives a median of only $3.70\%$ of the communication power, while
accounting for a median $46.03\%$ of the effective communication rate.
The supplementary stream benefits from decoding after sensing and echo
cancellation, but contributes \gls{ifi} while sensing is performed.
The optimized allocation therefore uses it selectively, obtaining a
substantial rate contribution from relatively little power while
limiting its sensing-stage interference. This asymmetric use of the two
streams helps explain how an interior message split can improve upon
either fixed processing order.

\subsection{\bc{Impact of the Sensing Requirement}}

\bc{Figure~\ref{fig:sensing_requirement} examines how the sensing requirement affects communication performance and the resulting \gls{rs}-inspired power allocation. The sweep uses $200$ common channel realizations with target \gls{rcs} $10$~dBsm and mean Doppler $3$~kHz. As defined in \eqref{eq:numerical_sensing_calibration}, smaller $\eta$ corresponds to a tighter sensing requirement. Results are reported separately at each requirement, without retaining higher-rate allocations obtained at tighter requirements.}

\bc{Figure~\ref{fig:sensing_requirement}(a) shows that relaxing the sensing requirement initially produces a substantial increase in communication performance. For the \gls{rs}-inspired design, the median effective \gls{se} increases from $6.91$~bps/Hz/subcarrier at $\eta=0.05$ to $8.86$~bps/Hz/subcarrier at $\eta=0.20$, while remaining above both fixed-order baselines throughout the sweep. It then decreases slightly to $8.82$~bps/Hz/subcarrier at $\eta=0.90$. Since relaxing the constraint enlarges the feasible set, this small decrease reflects the returned numerical solutions rather than a decrease in the globally optimal value.}

\begin{figure}[!htbp]
\vspace{-0.3cm}
\centering
\includegraphics[width=\columnwidth]{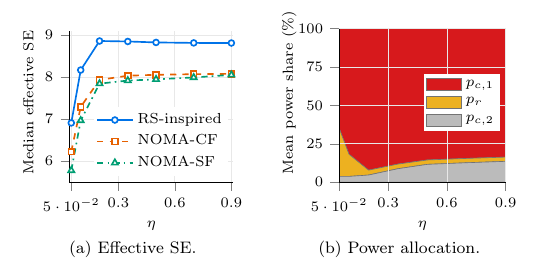}
\vspace{-0.3cm}
\caption{\bc{Sensing-requirement sweep: (a) median effective
\gls{se} (bps/Hz/subcarrier); (b) mean
\gls{rs}-inspired power shares.}}
\label{fig:sensing_requirement}
\end{figure}

\bc{The corresponding power allocations in Fig.~\ref{fig:sensing_requirement}(b) show how the power split changes with the sensing requirement. As $\eta$ increases from $0.05$ to $0.20$, the mean sensing-power share falls from $31.28\%$ to $3.18\%$, while the robust-stream share rises from $65.03\%$ to $92.23\%$. The supplementary stream remains active, with its mean share increasing from $3.69\%$ to $4.60\%$. As $\eta$ increases further to $0.90$, the supplementary-stream share grows to $13.52\%$, while the robust and sensing shares decrease to $83.73\%$ and $2.75\%$, respectively. Thus, tightening the sensing requirement primarily requires additional dedicated sensing power, whereas further relaxation changes the communication-stream split with little variation in effective \gls{se}.}
\subsection{\bc{Disturbance Management across Processing Stages}}

Figure~\ref{fig:mismatch_decomposition} examines how the three
architectures distribute interference across the sensing and
communication processing stages at mean Doppler $3$~kHz, target
\gls{rcs} $10$~dBsm, and $\eta=0.10$. The disturbance powers are
normalized by the noise power before averaging. Panel~(a) compares the
communication-induced \gls{ifi} and radar self-\gls{ici} present during
sensing, while panel~(b) decomposes the supplementary-decoding
disturbance into unresolved diagonal echo uncertainty, off-diagonal
echo \gls{ici}, and reconstruction mismatch. \gls{noma}-CF is absent
from panel~(b) because it has no supplementary stream.

\begin{figure}[!htbp]
\centering
\includegraphics[width=\columnwidth]{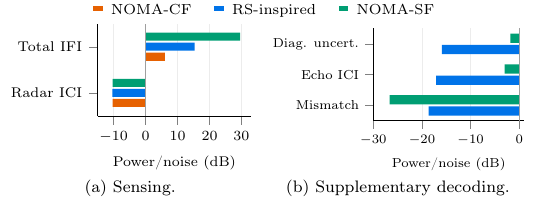}
\caption{\bc{Mean noise-normalized disturbance powers, expressed
in dB, for $200$ realizations at $\bar{\nu}=3$~kHz,
target \gls{rcs} $10$~dBsm, and $\eta=0.10$.
The noise reference is $0$~dB.}}
\label{fig:mismatch_decomposition}
\end{figure}

At the sensing stage, \gls{noma}-CF has the lowest \gls{ifi}, at
approximately $6.0$~dB above the noise power, since the communication
message is decoded before sensing. For \gls{noma}-SF, the corresponding
\gls{ifi} rises to approximately $29.6$~dB because the communication
signal remains present throughout sensing. The \gls{rs}-inspired design
lies between these endpoints at approximately $15.3$~dB, reflecting
the removal of the robust stream before sensing while retaining the
supplementary stream. Radar self-\gls{ici} remains similar across the
three architectures at approximately $-10.4$~dB relative to noise.

The complementary effect appears at supplementary decoding. Relative
to \gls{noma}-SF, the \gls{rs}-inspired design reduces both the
unresolved diagonal echo uncertainty and the off-diagonal echo
\gls{ici} by approximately $14.1$~dB, corresponding to a factor of
about $25.6$ in power. Its reconstruction mismatch is higher because
the reconstructed echo is also used to cancel the previously decoded
robust and sensing components, but remains approximately $18.6$~dB
below the noise power. Overall, the three architectures manage
interference differently across the processing stages: communication-first processing limits sensing-stage \gls{ifi}, sensing-first
processing shifts the communication message to the post-sensing stage,
while the \gls{rs}-inspired design distributes the message across both
stages to jointly manage sensing-stage \gls{ifi} and Doppler-induced \gls{ici}.

% ========================= CONCLUSION ====================================
\section{Conclusion}
\label{sec:conclusion}
% ================================================================================
This paper investigated an \gls{rs}-inspired uplink bistatic
\gls{ofdm}-\gls{isac} framework for jointly managing \gls{ifi} and
echo-induced \gls{ici}. By splitting the communication message across
pre- and post-sensing decoding stages, the proposed receiver generalizes
the fixed \gls{noma} communication-first and sensing-first processing orders, while
the sensing-constrained power allocation couples the sensing-error
surrogate to supplementary-stream decoding through modeled
echo-reconstruction mismatch. A tractable solution based on \gls{lmi}
reformulation, convex surrogate updates, and fractional programming was
developed. Numerical results show strict effective-\gls{se} gains over
the optimized fixed-order endpoints, with the message split adapting to
the echo-to-direct-path balance, Doppler-induced \gls{ici}, and sensing
requirement, including a transition toward the communication-first
endpoint under severe Doppler. The disturbance analysis further
illustrates how message splitting manages interference across processing
stages by balancing sensing-stage \gls{ifi} against post-sensing echo
\gls{ici}. Future work will consider empirical delay--Doppler estimation,
coded-link evaluation, imperfect direct-path \gls{csi}, and multi-target and multi-antenna extensions.
\end{anup}

% ========================= REFERENCES ====================================
\vspace{-1mm}
\bibliographystyle{IEEEtran}
\bibliography{refs}

@ARTICLE{Dai2026,
  author={Dai, Qianglong and Zeng, Yong and Wang, Huizhi and You, Changsheng and Zhou, Chao and Cheng, Hongqiang and Xu, Xiaoli and Jin, Shi and Lee Swindlehurst, A. and Eldar, Yonina C. and Schober, Robert and Zhang, Rui and You, Xiaohu},
  journal={IEEE Communications Surveys $\&$ Tutorials}, 
  title={A Tutorial on MIMO-OFDM ISAC: From Far-Field to Near-Field}, 
  year={2026},
  volume={28},
  number={},
  pages={4319-4358},
  doi={10.1109/COMST.2025.3650568}}

@ARTICLE{Moose1994,
  author={Moose, P.H.},
  journal={IEEE Trans. Commun.}, 
  title={{A technique for orthogonal frequency division multiplexing frequency offset correction}}, 
  year={1994},
  volume={42},
  number={10},
  pages={2908-2914},
  doi={10.1109/26.328961}}

@ARTICLE{Shen2018,
  author={Shen, Kaiming and Yu, Wei},
  journal={IEEE Trans. Signal Process.}, 
  title={{Fractional Programming for Communication Systems—Part I: Power Control and Beamforming}}, 
  year={2018},
  volume={66},
  number={10},
  pages={2616-2630},
  doi={10.1109/TSP.2018.2812733}}

@ARTICLE{Joshi2009,
  author={Joshi, Siddharth and Boyd, Stephen},
  journal={IEEE Trans. Signal Process.}, 
  title={{Sensor Selection via Convex Optimization}}, 
  year={2009},
  volume={57},
  number={2},
  pages={451-462},
  doi={10.1109/TSP.2008.2007095}}

@ARTICLE{Brunner2025,
  author={Brunner \textit{et al.}, David},
  journal={IEEE Trans. Microw. Theory Techn.}, 
  title={{Bistatic OFDM-Based ISAC With Over-the-Air Synchronization: System Concept and Performance Analysis}}, 
  year={2025},
  volume={73},
  number={5},
  pages={3016-3029},
  doi={10.1109/TMTT.2024.3487295}}

@techreport{3gpp_38901,
  author      = {3GPP},
  title       = {{Study on channel model for frequencies from 0.5 to 100 GHz}},
  institution = {3rd Generation Partnership Project (3GPP)},
  type        = {Technical Report (TR)},
  number      = {38.901 V19.1.0},
  month       = sep,
  year        = {2025},
  note        = {Release 19},
}

@ARTICLE{Garry2017,
  author={Garry \textit{et al.}, Joseph Landon},
  journal={IEEE Trans. Geosci. Remote Sens.}, 
  title={{Evaluation of Direct Signal Suppression for Passive Radar}}, 
  year={2017},
  volume={55},
  number={7},
  pages={3786-3799},
  doi={10.1109/TGRS.2017.2680321}}

@INPROCEEDINGS{Chen2020,
  author={Chen \textit{et al.}, Tingting},
  booktitle={Proc. IEEE Global Commun. Conf. (GLOBECOM)}, 
  title={{Location-Based Timing Advance Estimation for 5G Integrated LEO Satellite Communications}}, 
  year={2020},
  volume={},
  number={},
  pages={1-6},
  doi={10.1109/GLOBECOM42002.2020.9322428}}

@ARTICLE{Sturm2011,
  author={Sturm, Christian and Wiesbeck, Werner},
  journal={Proc. IEEE}, 
  title={{Waveform Design and Signal Processing Aspects for Fusion of Wireless Communications and Radar Sensing}}, 
  year={2011},
  volume={99},
  number={7},
  pages={1236-1259},
  doi={10.1109/JPROC.2011.2131110}}

@article{Liu2017,
title = {{Multiobjective optimal waveform design for OFDM integrated radar and communication systems}},
journal = {Signal Process.},
volume = {141},
pages = {331-342},
year = {2017},
issn = {0165-1684},
doi = {https://doi.org/10.1016/j.sigpro.2017.06.026},
author = {Liu \textit{et al.}, Yongjun},
}

@ARTICLE{Zhang2020,
  author={Zhang \textit{et al.}, Fuqiang},
  journal={IEEE Trans. Signal Process.}, 
  title={{Joint Range and Velocity Estimation With Intrapulse and Intersubcarrier Doppler Effects for OFDM-Based RadCom Systems}}, 
  year={2020},
  volume={68},
  number={},
  pages={662-675},
  doi={10.1109/TSP.2020.2965820}}

@book{Willis2004,
author = {Nicholas J. Willis},
title = {{Bistatic Radar, Chapter 25}},
publisher = {The Institution of Engineering and Technology},
year = {2004},
doi = {10.1049/SBRA003E},
edition   = {2nd},
}

@ARTICLE{Schniter2004,
  author={Schniter, P.},
  journal={IEEE Trans. Signal Process.}, 
  title={{Low-complexity equalization of OFDM in doubly selective channels}}, 
  year={2004},
  volume={52},
  number={4},
  pages={1002-1011},
  doi={10.1109/TSP.2004.823503}}

@ARTICLE{Sahin2023,
  author={Sahin \textit{et al.}, Mehmet Mert},
  journal={IEEE Commun. Lett.}, 
  title={{Multicarrier Rate-Splitting Multiple Access: Superiority of OFDM-RSMA Over OFDMA and OFDM-NOMA}}, 
  year={2023},
  volume={27},
  number={11},
  doi={10.1109/LCOMM.2023.3310937}}

@ARTICLE{Mishra2025,
  author={Mishra \textit{et al.}, Anup},
  journal={IEEE Commun. Lett.}, 
  title={{Coexistence of Radar and Communication with Rate-Splitting Wireless Access}}, 
  year={2025},
  volume={},
  number={},
  pages={1-1},
  doi={10.1109/LCOMM.2025.3625475}}

@INPROCEEDINGS{Tapio2024,
  author={Tapio \textit{et al.}, Visa},
  booktitle={Proc. IEEE 4th Int. Symp. Joint Commun. Sens. (JC$\&$S)}, 
  title={{Bi-Static Sensing with 5G NR Physical Uplink Shared Channel Transmission}}, 
  year={2024},
  volume={},
  number={},
  pages={1-6},
  doi={10.1109/JCS61227.2024.10646307}}

@ARTICLE{Sahin2025,
  author={Sahin \textit{et al.}, Mehmet Mert},
  journal={IEEE Trans. Commun.}, 
  title={{OFDM-RSMA: Robust Transmission Under Inter-Carrier Interference}}, 
  year={2025},
  volume={73},
  number={7},
  doi={10.1109/TCOMM.2024.3395713}}

@ARTICLE{Park2025bistatic,
  author={Park \textit{et al.}, Juha},
  journal={IEEE Trans. Aerosp. Electron. Syst.}, 
  title={{A Bistatic ISAC Framework for LEO Satellite Systems: A Rate-Splitting Approach}}, 
  year={2025},
  volume={},
  number={},
  pages={1-19},
  doi={10.1109/TAES.2025.3603500}}

@ARTICLE{Chiriyath2016,
  author={Chiriyath \textit{et al.}, Alex R.},
  journal={IEEE Trans. Signal Process.}, 
  title={{Inner Bounds on Performance of Radar and Communications Co-Existence}}, 
  year={2016},
  volume={64},
  number={2},
  pages={464-474},
  doi={10.1109/TSP.2015.2483485}}

@ARTICLE{Zhang2023SemiISAC,
  author={Zhang \textit{et al.}, Chao},
  journal={IEEE Trans. Commun.}, 
  title={{Semi-Integrated-Sensing-and-Communication (Semi-ISaC): From OMA to NOMA}}, 
  year={2023},
  volume={71},
  number={4},
  pages={1878-1893},
  doi={10.1109/TCOMM.2023.3241940}}

@INPROCEEDINGS{Wang2023,
author={Wang \textit{et al.}, Bowen},
booktitle={Proc. Int. Symp. Wireless Pers. Multimedia Commun. (WPMC)},
title={{{Cramer-Rao} Lower Bound Analysis for {OTFS} and {OFDM} Modulation Systems}},
year={2023},
pages={1--6},
}

@ARTICLE{Mishra2022,
  author  = {Mishra \textit{et al.}, Anup},
  journal = {IEEE Trans. Commun.}, 
  title   = {{Rate-Splitting Multiple Access for Downlink Multiuser {MIMO}: Precoder Optimization and {PHY}-Layer Design}}, 
  year    = {2022},
  volume  = {70},
  number  = {2},
  pages   = {874--890},
  doi     = {10.1109/TCOMM.2021.3138437}
}

@ARTICLE{Medard2000,
  author  = {Medard, M.},
  journal = {IEEE Trans. Inf. Theory}, 
  title   = {{The effect upon channel capacity in wireless communications of perfect and imperfect knowledge of the channel}}, 
  year    = {2000},
  volume  = {46},
  number  = {3},
  pages   = {933--946},
  doi     = {10.1109/18.841172}
}

@ARTICLE{Longfei2022a,
  author={Yin \textit{et al.}, Longfei},
  journal={IEEE Commun. Lett.}, 
  title={{Rate-Splitting Multiple Access for 6G—Part II: Interplay With Integrated Sensing and Communications}}, 
  year={2022},
  volume={26},
  number={10},
  pages={2237-2241},
  doi={10.1109/LCOMM.2022.3192032}}

@ARTICLE{Xu2021,
  author={Xu \textit{et al.}, Chengcheng},
  journal={IEEE J. Sel. Topics Signal Process.}, 
  title={{Rate-Splitting Multiple Access for Multi-Antenna Joint Radar and Communications}}, 
  year={2021},
  volume={15},
  number={6},
  pages={1332-1347},
  doi={10.1109/JSTSP.2021.3110312}}

@ARTICLE{FanLiu2022a,
  author={Liu \textit{et al.}, Fan},
  journal={IEEE J. Sel. Areas Commun.}, 
  title={{Integrated Sensing and Communications: Toward Dual-Functional Wireless Networks for 6G and Beyond}}, 
  year={2022},
  volume={40},
  number={6},
  pages={1728-1767},
  doi={10.1109/JSAC.2022.3156632}}

@ARTICLE{FanLiu2020,
  author={Liu \textit{et al.}, Fan},
  journal={IEEE Trans. Commun.}, 
  title={{Joint Radar and Communication Design: Applications, State-of-the-Art, and the Road Ahead}}, 
  year={2020},
  volume={68},
  number={6},
  pages={3834-3862},
  doi={10.1109/TCOMM.2020.2973976}}

@ARTICLE{Mishra@tutorial,
  author={Mishra \textit{et al.}, Anup},
  journal={IEEE Commun. Lett.}, 
  title={{Rate-Splitting Multiple Access for 6G—Part I: Principles, Applications and Future Works}}, 
  year={2022},
  volume={26},
  number={10},
  pages={2232-2236},
  doi={10.1109/LCOMM.2022.3192012}}

@ARTICLE{Yuanwei_NOMA_ISaC,
  author={Mu \textit{et al.}, Xidong},
  journal={IEEE Wireless Commun.}, 
  title={{{NOMA} for Integrating Sensing and Communications Toward {6G}: A Multiple Access Perspective}}, 
  year={2024},
  volume={31},
  number={3},
  pages={316-323},
  doi={10.1109/MWC.015.2200559}}

@ARTICLE{Yuanwei_Uplink_ISaC1,
  author={Ouyang \textit{et al.}, Chongjun},
  journal={IEEE Commun. Lett.}, 
  title={{On the Performance of Uplink {ISAC} Systems}}, 
  year={2022},
  volume={26},
  number={8},
  pages={1769-1773},
  doi={10.1109/LCOMM.2022.3178193}}

@article{mishra2025temporal,
  author={Mishra, Anup and De Souza, João Henrique Inacio and Popovski, Petar},
  journal={IEEE Trans. Commun.}, 
  title={Temporal Windows of Integration for Multisensory Wireless Systems as Enablers of Physical AI}, 
  year={2026},
  volume={},
  number={},
  pages={1-1},
  doi={10.1109/TCOMM.2026.3729724}}

@article{chen2025interference,
  title={{Interference Management for Integrated Sensing and Communications: A Multiple Access Perspective}},
  author={Chen \textit{et al.}, Kexin},
  journal={arXiv preprint arXiv:2509.02352},
  year={2025}
}

@ARTICLE{Wei_Survey,
  author={Wei \textit{et al.}, Zhiqing},
  journal={IEEE Internet Things J.}, 
  title={{Integrated Sensing and Communication Signals Toward 5G-A and 6G: A Survey}}, 
  year={2023},
  volume={10},
  number={13},
  pages={11068-11092},
  doi={10.1109/JIOT.2023.3235618}}

@INPROCEEDINGS{Li_Yiheng,
  author={Li \textit{et al.}, Yiheng},
  booktitle={Proc. IEEE/CIC Int. Conf. Commun. China (ICCC)}, 
  title={{Performance Analysis of Uplink Joint Communication and Sensing System}}, 
  year={2022},
  volume={},
  number={},
  pages={588-593},
  doi={10.1109/ICCC55456.2022.9880826}}

@article{Yu2025UplinkISACReceiver,
  title={{A framework for uplink ISAC receiver designs: performance analysis and algorithm development}},
  author={Yu \textit{et al.}, Zhiyuan},
  journal={arXiv:2503.02647},
  year={2025}
}

@INPROCEEDINGS{Hu2023UplinkRSMAISAC,
  author={Hu \textit{et al.}, Chao},
  booktitle={Proc. IEEE Wireless Commun. Netw. Conf. (WCNC)}, 
  title={{Joint Transmit and Receive Beamforming Design for Uplink RSMA Enabled Integrated Sensing and Communication Systems}}, 
  year={2023},
  volume={},
  number={},
  pages={1-6},
  doi={10.1109/WCNC55385.2023.10119024}}

@INPROCEEDINGS{Wang2022NOMAInspired,
  author={Wang \textit{et al.}, Zhaolin},
  booktitle={Proc. IEEE Int. Conf. Commun. (ICC)}, 
  title={{NOMA Inspired Interference Cancellation for Integrated Sensing and Communication}}, 
  year={2022},
  volume={},
  number={},
  pages={3154-3159},
  doi={10.1109/ICC45855.2022.9839026}}

@misc{Mishra2026RSInspiredUplink,
      title={Rate-Splitting-Inspired Uplink ISAC: A Rate-Region Analysis}, 
      author={Anup Mishra and Israel Leyva-Mayorga and Petar Popovski},
      year={2026},
      eprint={2606.07091},
      archivePrefix={arXiv},
      primaryClass={eess.SP},
      url={https://arxiv.org/abs/2606.07091}, 
}

@ARTICLE{XuZhang2024PriorISAC,
  author={Xu, Chan and Zhang, Shuowen},
  journal={IEEE Journal on Selected Areas in Communications}, 
  title={MIMO Integrated Sensing and Communication Exploiting Prior Information}, 
  year={2024},
  volume={42},
  number={9},
  pages={2306-2321},
  doi={10.1109/JSAC.2024.3413972}}
\end{document}